\documentclass[a4paper,11pt]{article}
\usepackage{jcappub} 
\usepackage[T1]{fontenc} 
\usepackage{ucs}
\usepackage[english]{babel}
\usepackage[T1]{fontenc}
\usepackage{amsmath}
\usepackage{amsfonts}
\usepackage{amssymb}
\usepackage{wrapfig}
\usepackage{graphicx,bbm,mathrsfs}  
\usepackage{epsfig}
\usepackage{rotating}
\usepackage{dcolumn}
\usepackage{bm}
\usepackage{float}
\usepackage[caption=false]{subfig}
\usepackage[normalem]{ulem}
\usepackage{soul,xcolor}
\usepackage{amsmath}
\usepackage{cancel}
\setstcolor{red}

\def\beq{\begin{equation}} 
\def\eeq{\end{equation}} 
\def\bea{\begin{eqnarray}}
\def\eea{\end{eqnarray}}
\newcommand{\ovl}[1]{\overline{#1}}
\newcommand{\wt}[1]{\widetilde{#1}}

\subheader{LPT-Orsay-17-28} 

\title{\boldmath $Z'$ portal to Chern-Simons Dark Matter}

\author[a]{Giorgio Arcadi}
\author[b]{Pradipta Ghosh}
\author[c]{Yann Mambrini}
\author[c]{Mathias Pierre}
\author[a]{Farinaldo S. Queiroz}

\affiliation[a]{Max Planck Institut f\"ur Kernphysik,\\ Saupfercheckweg 1, D-69117 Heidelberg, Germany}
\affiliation[b]{Department of Physics, Vidyasagar College,\\ 39, Sankar Ghosh Lane, Kolkata 700006, India}
\affiliation[c]{Laboratoire de Physique Th\'eorique, CNRS, Univ. Paris-Sud, 
Universit\'e  Paris-Saclay,\\ 91405 Orsay, France}

\emailAdd{arcadi@mpi-hd.mpg.de}
\emailAdd{tphyspg@gmail.com}
\emailAdd{yann.mambrini@th.u-psud.fr}
\emailAdd{mathias.pierre@th.u-psud.fr}
\emailAdd{queiroz@mpi-hd.mpg.de}

\abstract{We study the phenomenological credibility of a vectorial dark matter,
coupled to a $Z'$ portal through Chern-Simons interaction. We scrutinize two possibilities of connecting a 
$Z'$ with the Standard Model: (1) through kinetic mixing and (2) from a second Chern-Simons interaction.
Both scenarios are characterized by suppressed nuclear recoil scatterings, rendering direct detection 
searches not promising. Indirect detection experiments, on the other hand, furnish complementary limits 
for TeV scale masses, specially with the CTA. Searches for mono-jet and dileptons signals at the LHC 
are important to partially probe the kinetic mixing setup. Finally we propose an UV completion of 
the Chern-Simons Dark Matter framework.}


\arxivnumber{1706.04198}

\begin{document}
\maketitle
\flushbottom

\setcounter{equation}{0}


\section{Introduction}
\label{sec:intro}

The nature of Dark Matter (DM) remains one of the most exciting and puzzling mysteries of science 
\cite{Bertone:2004pz,Bertone:2010zza,Profumo:2013yn,Strigari:2013iaa,Bertone:2016nfn} till date. In 
the context of particle physics it is often assumed that one particle species can account for the 
entire DM abundance, which is $\sim$ 27\% of the budget of the Universe as indicated by PLANCK \cite{Ade:2015xua}. 
Such a sizable amount of {\it{hitherto undetected matter}} has diverse and intricate consequences for 
various ongoing and near future direct and indirect DM detection experiments \cite{Angle:2007uj,
Angle:2008we,Essig:2012yx,Aprile:2013doa,Aprile:2014eoa,Akerib:2015rjg,Agnes:2015ftt,Amole:2015pla,
Angloher:2015ewa,Hehn:2016nll,Akerib:2016lao,Aprile:2016swn,Amole:2016pye,Amole:2017dex}.

The absence of conclusive signals of new physics beyond the Standard Model (SM) concerning the DM leaves a 
handful of open debate, e.g., what are the particle characteristics of the DM or how they are interacting with the SM particles.
A popular solution for the DM puzzle is represented by the Weakly Interacting Massive Particles (WIMPs). In 
the simplest realization these new particle states, typically scalars or fermions, assumed to be singlet 
under the SM gauge group, feature interactions with the SM states mediated by the $Z$ or the SM-Higgs 
boson. Similar theoretical frameworks have also been addressed considering a 
vectorial DM~\cite{Hambye:2008bq,DiazCruz:2010dc,Mizukoshi:2010ky,Bhattacharya:2011tr,Farzan:2012hh,
Baek:2012se,Carone:2013wla,Chen:2014cbt,Graham:2015rva,Chen:2015dea,DiFranzo:2015nli,Bambhaniya:2016cpr,
Barman:2017yzr}. Unfortunately, such simple models, except for a very few exceptions, are critically 
challenged by the existing and expected upcoming DM searches from the direct, indirect and collider probes.
A detailed overview of these frameworks are addressed recently in 
ref.~\cite{Arcadi:2017kky} (see also ref.~\cite{Escudero:2016gzx}).

Considering apparent limitations of these simplest realizations, as extensively addressed in ref.~\cite{Arcadi:2017kky}, a natural extension of the above setup is provided by the so called ``simplified models'' (see ref.~\cite{Abdallah:2015ter})
in which the 
SM mediators are replaced by a new ``dark mediator'', of various spin assignations. This dark mediator 
can be connected to the SM particles
by various ways and hence, relates the otherwise secluded DM
to the SM particles. However, to ensure elucidate
predictions, it is nevertheless necessary to investigate 
theoretical competence of these simplified setups~\cite{Kahlhoefer:2015bea,Englert:2016joy,Bell:2016uhg,Goncalves:2016iyg,Bell:2016ekl}, i.e.,
for example, whether they have consistent unitarity behaviours, reasonable Ultra-Violet (UV) completion and 
how they can be embedded into unified theory frameworks.

The case of spin-1 mediators, among these setups, perhaps deserves serious attentions as phenomenological 
study of these frameworks reveals intricate complementary aspects of DM searches with 
relevant collider observations (see ref.~\cite{Arcadi:2017kky} for a thorough discussion). 
An intriguing origin of such spin-1 mediator(s)
can arise as gauge boson(s) of some beyond the SM (BSM) gauge group(s),
Abelian or non-Abelian, that simultaneously assigns non-zero
gauge charge(s) also for the DM candidate. A spin-1 mediator,
maintaining gauge invariance and renormalizability,
can couple to the SM Electro-Weak (EW) gauge boson, and thus, subsequently to other SM particles, in a few 
different ways,\footnote{A BSM spin-1 mediator can couple
to the SM EW gauge bosons also via the well-known spontaneous symmetry breaking and Higgs mechanism, provided that the SM-Higgs
doublet has non-zero charges under the BSM gauge groups. Spontaneous symmetry breaking in the BSM sector is triggered 
with new SM singlet scalar which may or may not mix
with the SM-Higgs. We do not consider this possibility in our analysis.} e.g., via a kinetic mixing\footnote{A kinetic mixing, from the principle of gauge invariance, is allowed only between the vector bosons of Abelian groups.}
\cite{Holdom:1985ag,delAguila:1988jz,delAguila:1995rb,Foot:1991kb} or using a Chern-Simons (CS) interaction~\cite{Dudas:2009uq,Mambrini:2009ad,Dudas:2012pb,Dudas:2013sia}. The former can either appear naturally
in Lagrangian preserving the gauge invariance and renormalizability of the SM \cite{Holdom:1985ag,delAguila:1995rb,Foot:1991kb} or can get generated after integrating
out the heavy fermionic degrees of freedom \cite{Holdom:1985ag,delAguila:1988jz} charged under both the SM and BSM gauge groups. The latter can also 
arise in an analogous way after integrating out such heavy degrees of freedom,
as extensively studied in ref.~\cite{Anastasopoulos:2006cz}. The presence of these new fermionic degrees
of freedom, if chiral under some representations, introduces new challenges to construct an anomaly free model framework.\footnote{One can always 
consider these new fermions to transform vector-like with respect to the SM gauge groups such that no new chiral anomalies appear in the SM.}
This goal is customarily achieved by arranging anomaly cancellation in the chiral sectors of the theory, by assigning specific couplings/charges
for the involved particle species, e.g., by considering distinct couplings between the SM chiral fermions, 
and possibly also the DM, with the BSM spin-1 mediator as discussed in ref.~\cite{Arcadi:2017atc} in the 
context of unified theories. Some classes of anomalies, like the triangle ones
involving Abelian, non-Abelian or a mixture of the two gauge groups can, alternatively be cured through the 
Green-Schwarz mechanism~\cite{Green:1984sg,Green:1984ed}.

In this work we aim to study phenomenological viability of models,
consisting of a vectorial DM candidate connected to the SM particles via a $Z'$-portal
~\cite{Arcadi:2013qia,Profumo:2013sca,Cline:2014dwa,Lebedev:2014bba,deSimone:2014pda,
Feng:2014eja,Buchmueller:2014yoa,Fairbairn:2014aqa,Allanach:2015gkd,Alves:2015mua,Alves:2015pea,
Kahlhoefer:2015bea,Undagoitia:2015gya,Ducu:2015fda,Fairbairn:2016iuf,Klasen:2016qux,Englert:2016joy,
Alves:2016cqf,Jacques:2016dqz,Altmannshofer:2016jzy,Alves:2016cqf,Okada:2016tci,Alves:2016fqe}. 
We consider specific frameworks that are inspired by Green-Schwarz mechanism~(see also refs.~\cite{Dudas:2009uq,Dudas:2012pb}), 
i.e., the connection between the DM and the $Z'$ is mediated by a CS interaction.
When considering theoretical arguments to account for the DM stability, for example an extended gauge structure, 
a vectorial DM candidate appears naturally and can be  identified as the cosmologically stable gauge boson of a 
new BSM symmetry group provided that this gauge boson is the lightest state charged under this extra symmetry group 
while the $Z'$ can arise from another
BSM Abelian or non-Abelian gauge groups. CS interactions with $\mathcal{O}(1)$ couplings are possible by 
considering heavy chiral fermionic content charged under these extra gauge symmetry groups in an anomaly-free setup. 
This represents an intriguing possibility for DM phenomenology as both the DM candidate and its coupling to the 
$Z^\prime$ mediator would be a consequence of some extended gauge structure. Confining within the framework of 
Abelian theory, we will explore two possibilities of how a $Z'$
can coordinate with the SM. These examples, as already mentioned, consider:
(1) gauge invariant renormalizable
kinetic mixing between the field strength of $Z'$ with $B_{\mu\nu}$, the SM hypercharge field strength
and, (2) a second CS interaction involving $Z'_\mu$
and the SM hypercharge vector $B_\mu$.
For both these scenarios we extensively investigate the impact of measured relic density \cite{Ade:2015xua} as well 
as the existing and anticipated sensitivity reaches
from Direct Detection (DD) and Indirect Detection (ID) experiments
on the associated model parameter spaces. In this setup we consider the vectorial DM candidate as a typical WIMP 
in a $\Lambda$CDM cosmology~\cite{Ade:2013zuv,Ade:2015xua} and we further assume that the so called "small scale 
controversies"~\cite{Klypin:1999uc,Moore:1999nt,Strigari:2010un,BoylanKolchin:2011de,Weinberg:2013aya} to be 
unrelated and independent of the DM nature and properties.\footnote{It has been recently pointed out that the 
baryonic feedback might alleviate existing tensions with N-body simulations (see e.g., 
refs.~\cite{Schaller:2014uwa,Cautun:2015dqa,Sawala:2015cdf,Tollet:2015gqa,Dutton:2015nvy}).}
We also explore relevant theoretical constraints like EW Precision Tests (EWPTs), UV completion etc. for these setups. Finally, 
for completeness, we also discuss the possible pertinent collider aspects of such models, e.g., 
invisible $Z$-decay width, mono-{\bf X}, dijets, dilepton searches, etc.

The paper is organized as follows: in the next two sections we will study phenomenologies of the two 
aforesaid scenarios relying on the low-energy effective Lagrangians without specifying theoretical issues 
like UV completion etc. Section \ref{sec:UVcompletion} will be dedicated for this task where we will
address the generation of a CS coupling as well as 
a kinetic mixing in a UV complete setup.
We will present the summary of our analyses and put our
concluding remarks in section \ref{sec:Conclusion}. Some useful
formulae like detail constructions of an anomaly free model are 
relegated to the appendices.

\section{Scenario-I: $Z'$-$Z$ interaction via kinetic mixing}
\label{sec:scenario1}

In this section we study the aforesaid type-I scenario when
the ``dark sector'', comprised of a vectorial DM $X_\mu$ and a spin-1
vector boson $\wt V_{\mu}$, is ``secluded'' from the visible sector,
i.e., there exists no direct coupling between this dark sector and the SM 
fermions.\footnote{The other possibility, i.e., the dark sector has direct 
couplings with the SM fermions is reviewed 
recently in ref.~\cite{Arcadi:2017kky}.} These $X_{\mu}$
and $\wt V_\mu$, for example, can appear as the gauge bosons of 
some BSM $U(1)_X$ and $U(1)_V$ groups and we consider a
CS interaction to connect them together.
As already discussed, a bridge between the dark sector and the SM now appears via a kinetic mixing of $\wt V_{\mu\nu}$ and
$B_{\mu\nu}$, the field strengths associated with BSM $U(1)_V$
and the SM $U(1)_Y$ gauge group, respectively. A similar kinetic mixing between $X_{\mu \nu} $ and $B_{\mu \nu}$,
being renormalizable and allowed by the SM gauge invariance,
should also be included in a general Lagrangian. However, we do not consider this possibility for the stability of the DM and 
postpone further discussion in this direction till section \ref{sec:UVcompletion}.

The relevant phenomenology of the said model can be described by the following low-energy effective Lagrangian:
\bea
\label{eq:starting_lagrangian}
 \mathcal{L}\supset &&-\frac{1}{4} {B}^{\mu \nu} {B}_{\mu \nu}-\frac{1}{4} X^{\mu \nu} X_{\mu \nu} -\frac{1}{4} \wt{V}^{\mu \nu} \wt{V}_{\mu \nu}  - \frac{\sin \delta}{2}\widetilde{V}^{\mu \nu}{B}_{\mu \nu} \nonumber\\
&& + \alpha_{\rm CS} \epsilon^{\mu \nu \rho \sigma}  X_{\mu} \wt{V}_{\nu} X_{\rho \sigma} +\dfrac{m_{V}^2}{2}\wt{V}^\mu \wt{V}_\mu+\dfrac{m_{X}^2}{2}X^\mu X_\mu,
\eea
here $\delta$ is the kinetic mixing parameter and $\alpha_{\rm CS}$
represents the effective coupling of CS operator.
$X_{\mu\nu}$ gives the field strength of $U(1)_X$ group
and $m_V,\,m_X$ represent mass terms of the mediator and the DM.
Thus, one gets a set of four free inputs, namely,
$\delta,\, \alpha_{\rm CS}$, $m_V$ and $m_X$ whose ranges
will be tested subsequently imposing a series of theoretical
and experimental constraints.

The presence of kinetic mixing in eq.~(\ref{eq:starting_lagrangian}) implies non-canonical kinetic
term for $B_{\mu\nu}$ and also for $\wt V_{\mu\nu}$. In order 
to generate diagonal kinetic terms in the physical or mass
basis one should invoke three different rotations
~\cite{Babu:1997st,Chun:2010ve,Mambrini:2010dq,Mambrini:2011dw}. 
The first 
rotation, involving then angle $\delta$, takes $B_{\mu},\,
\wt V_{\mu}$ to a basis (say $B^\text{int}_{\mu},\,
\wt V^\text{int}_{\mu}$) with diagonal kinetic terms. The
second rotation, after EW symmetry breaking (EWSB), via angle $\theta_{\wt W}$, takes this $B^\text{int}_{\mu}$
together with $W^3_{\mu}$ to the intermediate $A_\mu,\,Z^\text{int}_\mu$ basis. Finally,
the third rotation through another angle $\phi$, leaving
the massless photon aside, takes $Z^\text{int}_\mu,\,\wt V^\text{int}_\mu$
to the $Z_\mu,\,Z'_\mu$ basis where $Z_\mu,\,Z'_\mu$
are associated with the physical $Z$ and $Z'$ boson. 
In summary, the initial $B_\mu,\, W^3_\mu,\, \wt V_\mu$
basis can be related to the physical $A_\mu,\,Z_\mu,\,Z'_\mu$
basis in the following way:

\begin{equation}
\label{eq:transformation}
\begin{bmatrix}
{B}_\mu \\ 
W_{3 \mu} \\
\wt{V}_{\mu}
\end{bmatrix}=\begin{bmatrix}
{c}_{\wt W} & -{s}_{\wt W} c_\phi + t_\delta s_\phi &-{s}_{\wt W} s_\phi - t_\delta c_\phi \\ 
{s}_{\wt W} & {c}_{\wt W} c_\phi & {c}_{\wt W} s_\phi \\ 
0 & -\dfrac{s_\phi}{c_\delta} & \dfrac{c_\phi}{c_\delta}
\end{bmatrix}
\begin{bmatrix}
A_\mu \\ 
Z_\mu \\
Z^\prime_\mu
\end{bmatrix},
\end{equation}
where $t_\delta,\,c_\delta,\,c_{\wt W},\,s_{\wt W} ,\,s_\phi,\,c_\phi \equiv \tan\delta,\,\cos\delta,\, \cos\theta_{\wt W},\,\sin\theta_{\wt W},\,\sin\phi,\,\cos\phi$ with:
\begin{equation}
\tan2 \phi=\dfrac{\wt{m}_Z^2 {s}_{\wt W} \sin 2\delta}{m_V^2-\wt{m}_{Z}^2(c_\delta^2 -{s^2}_{\wt W} s_\delta^2)},
\end{equation}
here $s_\delta\equiv \sin\delta$. The quantities $\theta_{\wt W},\,
\wt m_Z$ do not represent the measured values of Weinberg angle
and $Z$-boson mass \cite{Olive:2016xmw} but are related to
them as will be explained later.
These quantities are obtained after the SM EWSB using a rotated $B_\mu$ field (a basis where off-diagonal mixing term of eq. (\ref{eq:starting_lagrangian}) between $B_\mu$ and $\widetilde V_\mu$ vanishes) and the $W^3_\mu$ field of the SM.
The masses of $Z^\prime$ and $Z$ are written as:
\beq
\label{eq:mzzpmass}
m_{Z^\prime,Z}^2=\dfrac{1}{2}\left[ \wt{m}_Z^2(1+{s^2}_{\wt W} t_\delta^2)+\dfrac{m_V^2}{c_\delta^2}    \pm \sqrt{(\wt{m}_Z^2(1+{s^2}_{\wt W} t_\delta^2)+\dfrac{m_V^2}{c_\delta^2})^2 - \dfrac{4}{c_\delta^2}\wt{m}_Z^2 m_V^2 } \right],
\eeq
which gives $m_Z \simeq \wt m_Z$ in the experimentally favoured limit $\delta \ll 1$, along with $m_{Z'} \simeq m_V$.

Notice that the transformation used in eq.~(\ref{eq:transformation}) is valid only if one of these two conditions is met:
\bea
\label{eq:masses}
 \frac{m_V^2}{\wt{m}_Z^2} &&\geq 1+2 s_{\wt{W}} \tan^2 \delta +2 \sqrt{s_{\wt{W}}^2 \tan^2 \delta \left(1+s_{\wt{W}}^2 \tan^2 \delta\right)}, \nonumber\\
 \frac{m_V^2}{\wt{m}_Z^2} &&\leq 1+2 s_{\wt{W}} \tan^2 \delta -2 \sqrt{s_{\wt{W}}^2 
 \tan^2 \delta \left(1+s_{\wt{W}}^2 \tan^2 \delta\right)}.
\eea

One should note that the transformation of eq.~(\ref{eq:transformation})
does not change the photon coupling~\cite{Babu:1997st}, implying the following identity:
\begin{equation}
\label{eq:stwswrelation}
s_{\wt W} c_{\wt W} \wt{m}_Z^2=s_W c_W m_Z^2=\frac{\pi \alpha(m_Z)}{\sqrt{2} G_F},
\end{equation}
where $s_W,\,c_W\equiv \sin\theta_W,\,\cos\theta_W$ are associated with the measured value of Weinberg angle 
$\theta_W$ \cite{Olive:2016xmw}. $\alpha(m_Z)$ is the fine structure constant at the energy scale $m_Z$ and $G_F$
represents the Fermi constant \cite{Olive:2016xmw}. 

One should also consider the invariance of $W$-boson mass
under the transformation of eq.~(\ref{eq:transformation}), i.e., $m_W^2=m^2_Z c_W^2=\wt{m}_Z^2 {c^2}_{\wt W}$ which 
allows us to express the $\rho$ 
parameter \cite{Olive:2016xmw} as:
\begin{equation}
\label{eq:rhoparam}
\rho = \frac{\wt{m}_Z^2 {c^2}_{\wt W}}{m^2_Z c^2_W},
\end{equation} 
with the experimental measured value given by $\rho-1=4^{+8}_{-4} \times 10^{-4}$~\cite{Olive:2016xmw}.
Further, from the EWPT one can consider a simple and conservative limit on 
$\delta$ as~\cite{Cassel:2009pu,Hook:2010tw,Kumar:2006gm,Chun:2010ve}:
\begin{equation}
\label{eq:EWPT}
\delta \lesssim \arctan \left[0.4 \left( \dfrac{m_{Z'}}{\text{TeV}} \right) \right].
\end{equation}

It is now apparent that one can use eq.~(\ref{eq:stwswrelation}),
eq.~(\ref{eq:rhoparam}) and eq.~(\ref{eq:EWPT})
to discard experimentally disfavoured values of the kinetic mixing parameter $\delta$.
Further constraints on $\delta$ can emerge from various other experimental observations.
The mixing among $B_\mu,\,W^3_\mu $ and $\wt V_\mu$ (see eq.~(\ref{eq:transformation})), 
couples the SM fermions and the DM with the $Z$ boson. The latter coupling implies
an enhancement of the invisible $Z$ decay width for $2 m_X < m_Z$.  Hence, the parameter $\delta$, along with $m_V$, 
will receive constraints from a plethora of different collider searches like dileptons $(pp\to Z'$ $\to e^+e^-$, $\mu^+\mu^-)$ \cite{
Aaboud:2016cth,Khachatryan:2016zqb,ATLAS:2017wce}, dijets $(pp\to Z'\to q \ovl q)$ 
\cite{Khachatryan:2015dcf,ATLAS:2015nsi,Sirunyan:2016iap,Aaboud:2017yvp}, mono-{\bf X} $(pp\to Z'+{\mathbf{X}},\, 
Z'\to {\rm DM~pairs})$ with ${\mathbf{X}}=q/g$
\cite{Sirunyan:2017hci}, $W$ \cite{Aaboud:2016qgg}, $Z$ \cite{Aaboud:2016qgg,CMS:2016hmx}, 
$\gamma$ \cite{Aaboud:2016uro,CMS:2016fnh,Aaboud:2017dor},
SM-Higgs \cite{ATLAS:2017uwx,ATLAS:2017pqx,Sirunyan:2017hnk}, etc., 
invisible $Z$ decay width~\cite{Olive:2016xmw} and a few others. The 
dijets and mono-{\bf X} searches also restrict the parameters $\alpha_{\rm CS},\,m_X$. 
Finally, the DM phenomenology, i.e., the correct relic density, DD and ID results will also put limits
on the parameters $\delta,\,\alpha_{\rm CS},\,m_V$ and $m_X$.
We note in passing that for numerical analyses we have traded the parameter 
$m_V$ with the physical mass $m_{Z'}$ using eq.~(\ref{eq:mzzpmass}).

In the mass basis, the Lagrangian relevant for our subsequent analysis is given by: 
\bea
\label{eq:mass_basis_lagrangian}
 \mathcal{L}_{Z/Z',SM}&&=\ovl f \gamma^\mu \left(g_{f_L}^{Z}P_L+g_{f_R}^{Z}P_R\right) f Z_\mu
 +\ovl f \gamma^\mu \left(g_{f_L}^{Z'}P_L+g_{f_R}^{Z'}P_R\right) f Z'_\mu  \nonumber\\
&& + g_W^Z [[W^+ W^-Z]]+g_W^{Z'} [[W^+ W^-Z']] +\dfrac{g_{hZZ}}{2} Z^\mu Z_\mu h + g_{hZZ'}  Z'_\mu Z^\mu h \nonumber\\
&& +\dfrac{g_{hZ'Z'}}{2}  Z'_\mu Z'^{\mu} h- \alpha_{\rm CS}\,\frac{s_\phi}{c_\delta}\,\epsilon^{\mu \nu \rho \sigma}\, X_\mu X_{\rho \sigma} Z_\nu+ \alpha_{\rm CS}\,\frac{c_\phi}{c_\delta}\,\epsilon^{\mu \nu \rho \sigma}\, X_\mu X_{\rho \sigma} Z^{'}_\nu,
\eea
here $P_{L(R)}=(1\pm\gamma_5)/2$ and
%
\bea
\label{eq:zzpsmfermion}
g^{Z}_{f_L}&&= g_Y Y_L (- s_W c_{\phi }+t_{\delta }s_{\phi })+g_W I_z (c_W c_\phi),\,\,\,\,
g^{Z}_{f_R}= g_Y Y_R (- s_W c_{\phi }+t_{\delta }s_{\phi }),\nonumber\\
g^{Z^\prime}_{f_L}&&= g_Y Y_L (- s_W s_{\phi }-t_{\delta }c_\phi)+g_W I_z (c_W s_\phi),\,\,\,\,
g^{Z^\prime}_{f_R}= g_Y Y_R (- s_W s_{\phi }-t_{\delta }c_\phi),
\eea
where $g_Y$ is the gauge coupling of $U(1)_Y$,
$I_z$ is the $3^{rd}$ component of the weak isospin, $Y_{L,R}$ are hypercharge of the left- and 
right- chiral fermions, $g_W$ is the $SU(2)_L$ gauge coupling.
\beq
\label{eq:zzpww}
g^{Z}_{W}=g_W c_W c_\phi,\,\,\, \qquad
g^{Z^\prime }_{W}=g_W c_W s_\phi,
\eeq
%
\bea
\label{eq:zzph}
g_{hZZ}&&=2 v_h \Big[ \frac{g_Y}{2} (- s_W c_{\phi }+t_{\delta }s_{\phi }) -\frac{g_W}{2}(c_W c_\phi) \Big]^2,\nonumber \\
g_{hZZ^\prime }&&=2  v_h \Big[ \frac{g_Y}{2} (- s_W s_{\phi }-t_{\delta }c_{\phi }) -\frac{g_W}{2}(c_W s_\phi) \Big]\Big[ \frac{g_Y}{2} (- s_W c_{\phi }+t_{\delta }s_{\phi }) -\frac{g_W}{2}(c_W c_\phi) \Big],\nonumber\\
g_{hZ^\prime Z^\prime }&&=2 v_h \Big[ \frac{g_Y}{2} (- s_W s_{\phi }-t_{\delta }c_{\phi }) -\frac{g_W}{2}(c_W s_\phi) \Big]^2,
\eea
here $v_h$ is the vacuum expectation value (VEV) of the SM-Higgs
doublet and a factor `$2$' appears in $ZZh,\,Z'Z'h$ vertices
due to the presence of identical particles.
For the 
$\alpha_{\rm CS}\,\frac{-s_\phi(c_\phi)}{c_\delta}$ $\epsilon^{\mu \nu \rho \sigma}$  
$X_\mu X_{\rho \sigma} Z_\nu(Z^{'}_\nu)$ term, 
considering the following momentum assignments
$X_{\mu}(p_a) X_\nu (p_b)$ $Z_{\rho}/Z'_\rho$ one can use
\beq
\label{eq:zzpdm}
g_{X}^Z =-2 \alpha_{\rm CS}\, 
\frac{s_\phi}{c_\delta}, \,\,\,\,
g^{Z^\prime}_X =2 \alpha_{\rm CS}\,
\frac{c_\phi}{c_\delta},
\eeq
for subsequent relevant analytical formulae like DM pair annihilation
cross-section etc. Once again the factor `$2$' appears
due to identical particles. Lastly
\bea
[[W^+ W^- Z(Z')]] &&\equiv i \left[W^+_{\mu\nu} W^{-\, \mu}Z(Z')^\nu-W^+_{\,\mu\nu} W^{-\,\mu}Z(Z')^\nu \right. \nonumber\\
&& \left.~~~~+\frac{1}{2}Z(Z')^{\mu\nu} \left(W^{+}_\mu W^{-}_\nu-W^{+}_\nu W^{-}_\mu\right)\right].
\eea

With these analytical expressions now we will discuss the DM phenomenology
of this framework in the next subsection.

\subsection{Dark Matter Phenomenology}

In this subsection we discuss the DM phenomenology in the light
of various constraints coming from the requirement of correct relic density
and consistency with the existing and/or upcoming DD and ID results. 
The DM is produced in the early Universe according to the 
WIMP paradigm. We will subsequently discuss
how these
observations can affect the accessibility of the chosen set-up
at the present or in the near future DM detection experiments, assuming projected search sensitivities.
Nevertheless, for the sake of completeness, we will also qualitatively discuss the 
complementary limits arising
from the EWPT, collider searches, etc. 
We start our discussion in the context of accommodating correct relic density
and successively will address the restrictions coming from direct
and indirect DM searches.

\subsubsection{Relic density}
\label{sssec:rdsc1}

It is well known that according to the WIMP paradigm a single particle
species can account for the correct DM relic density $\Omega_X h^2$
$\propto 1/\langle \sigma v \rangle$, where $\langle \sigma v \rangle$
represents the thermally averaged DM pair annihilation cross-section into the SM states. The experimentally 
hinted value $\Omega_X h^2 \approx 0.12$~\cite{Ade:2015xua} predicts $\langle \sigma v 
\rangle$, $\sim \mathcal{O}$ $(10^{-26}\,{\mbox{cm}}^3\,{\mbox{s}}^{-1})$.
In the chosen framework (see Section~\ref{sec:scenario1}), the DM pair can annihilate into the SM fermions 
$(\ovl f f)$ and $W^+W^-$, as well as in $Zh$ and $Z'h$ final states, through s-channel exchange of the $Z/Z'$ boson, 
and into $ZZ,\,ZZ'$ and $Z'Z'$ final states, through t-channel exchange of a DM state. Approximate analytical 
expressions of the corresponding annihilation cross-sections are obtained through the expansion $\langle 
\sigma v \rangle \simeq a +b \dfrac{1}{x}+...$ with $x=\dfrac{m_X}{T}$~\cite{Gondolo:1990dk,Jungman:1995df}, $T$
being the temperature. We remind, however, that the aforesaid approximation is not reliable over the entire 
parameter space~\cite{Griest:1990kh}, e.g., the pole regions $m_X \sim \dfrac{m_{Z,\,Z'}}{2}$. Hence, all results presented 
in this article are obtained through full numerical computations using the package 
micrOMEGAs~\cite{Belanger:2006is,Belanger:2008sj,Belanger:2013oya},
after implementing the model in FeynRules \cite{Christensen:2008py,Alloul:2013bka}.
The DM annihilation cross-sections in the possible final states are
given as:
\begin{itemize}
\item $XX \to \ovl{f}f:$
\bea
\label{eq:XXff}
\langle \sigma v\rangle_{\bar{f}f} &&\simeq  \frac{n_c\,m_f^2}{3 \pi} \sqrt{1-\frac{m_f^2}{m_X^2}} {\left(\frac{\left(g_{f_L}^Z-g_{f_R}^Z\right) g_X^Z}{m_Z^2}+\frac{\left(g_{f_L}^{Z'}-g_{f_R}^{Z'}\right) g_X^{Z'}}{m_{Z'}^2}\right)}^2\dfrac{1}{x}\nonumber\\
&&+\frac{n_c}{108 \pi}\left[ {\left( \frac{g^Z_X g^Z_{f_L}}{m^2_Z} + \frac{g^{Z'}_X g^{Z'}_{f_L}}{m^2_{Z'}} \right)}^2 + {\left( \frac{g^Z_X g^Z_{f_R}}{m^2_Z} + \frac{g^{Z'}_X g^{Z'}_{f_R}}{m^2_{Z'}} \right)}^2 \right ] \frac{1}{x^2}
\eea
where we have written not only the leading $1/x$ (p-wave) term but also
the second order $1/x^2$ (d-wave) one which appears to be
the dominant one since the first one is suppressed by the square
of SM fermion mass $m^2_f$. The parameter $n_c$ denotes colour factor
with a value of $3\,(1)$ for quarks (leptons).
\item $XX\to W^+W^-:$
\beq
\label{eq:XXWW}
\langle \sigma v\rangle_{W^+W^-}\simeq \frac{5}{9 \pi m_W^4}{\left(\frac{g_X^Z g_W^Z m_X}{4}-\frac{g_X^{Z'} g_W^{Z'} m_X^3}{m_{Z'}^2}\right)}^2\dfrac{1}{x^2},
~~{\rm for~} m_{W,\,Z} \ll m_X \ll m_{Z^\prime}.
\eeq
%
\item $XX\to Zh:$
\beq
\label{eq:XXZh}
\langle \sigma v\rangle_{Zh}\simeq \frac{2}{3\pi}\frac{m_X^6}{m_Z^6}{\left(\frac{g_X^Z g_{hZZ}}{4 m_X^2}-\frac{g_X^{Z'} g_{hZZ'}}{m_{Z'}^2}\right)}^2\dfrac{1}{x},
~~{\rm for~}m_Z \sim m_h \ll m_X \ll m_{Z^\prime}.
\eeq
%
\item $XX\to Z^\prime h:$
\beq
\label{eq:XXzph}
\langle \sigma v\rangle_{Z^\prime h}\simeq \frac{1}{24 \pi}\frac{m_X^2}{m_{Z'}^6}{\left(g_{hZZ'} g_X^Z+g_{hZ'Z'} g_X^{Z'}\right)}^2 \dfrac{1}{x},
~~{\rm for~}m_Z \sim m_h  \ll m_{Z^\prime} \ll m_X.
\eeq
%
\item $XX\to ZZ:$
%
\beq
\label{eq:XXZZ}
\langle \sigma v\rangle_{ZZ}\simeq\frac{8 \alpha_{\rm CS} ^4 \delta ^4 m_X^2 m_Z^4 s_W^4}{9 \pi  m_{Z'}^8},
~~{\rm for~} m_Z \ll m_X  \ll m_{Z^\prime}.
\eeq
%
\item $XX\to Z^\prime Z^\prime:$ 
%
\beq
\label{eq:XXzpzp}
\langle \sigma v\rangle_{Z^\prime Z^\prime}\simeq \frac{8 \alpha_{\rm CS} ^4 m_X^2}{9 \pi  m_{Z'}^4},
~~{\rm for~} m_Z   \ll m_{Z^\prime}  \ll m_X.
\eeq
%
\item $XX\to Z^\prime Z:$ 
\beq
\label{eq:XXzpz}
\langle \sigma v\rangle_{Z^\prime Z}\simeq \frac{16 \alpha_{\rm CS} ^4 \delta ^2 m_X^2 m_Z^2 s_W^2}{9 \pi  m_{Z'}^6},
~~{\rm for~} m_Z   \ll m_{Z^\prime}  \ll m_X.
\eeq
\end{itemize}

As evident from the above expressions that annihilation channels originated by s-channel exchange of the $Z/Z'$ feature a 
suppressed annihilation cross-section, at least p-wave or even d-wave in the cases of $W^+ W^-$ and $\ovl f f$ final states  
(see eq.~(\ref{eq:XXWW})
and eq.~(\ref{eq:XXff})). For the latter a p-wave contribution is also present but it is helicity suppressed. 
The t-channel induced annihilations feature, instead, a s-wave cross-section. The cross-section into $ZZ$ pairs 
(see eq.~(\ref{eq:XXZZ})) is anyway suppressed by an higher power of the kinetic mixing parameter $\delta$, with 
respect to the other annihilation channels.     

\begin{figure}[t!]
\centering
\includegraphics[width=7.5cm]{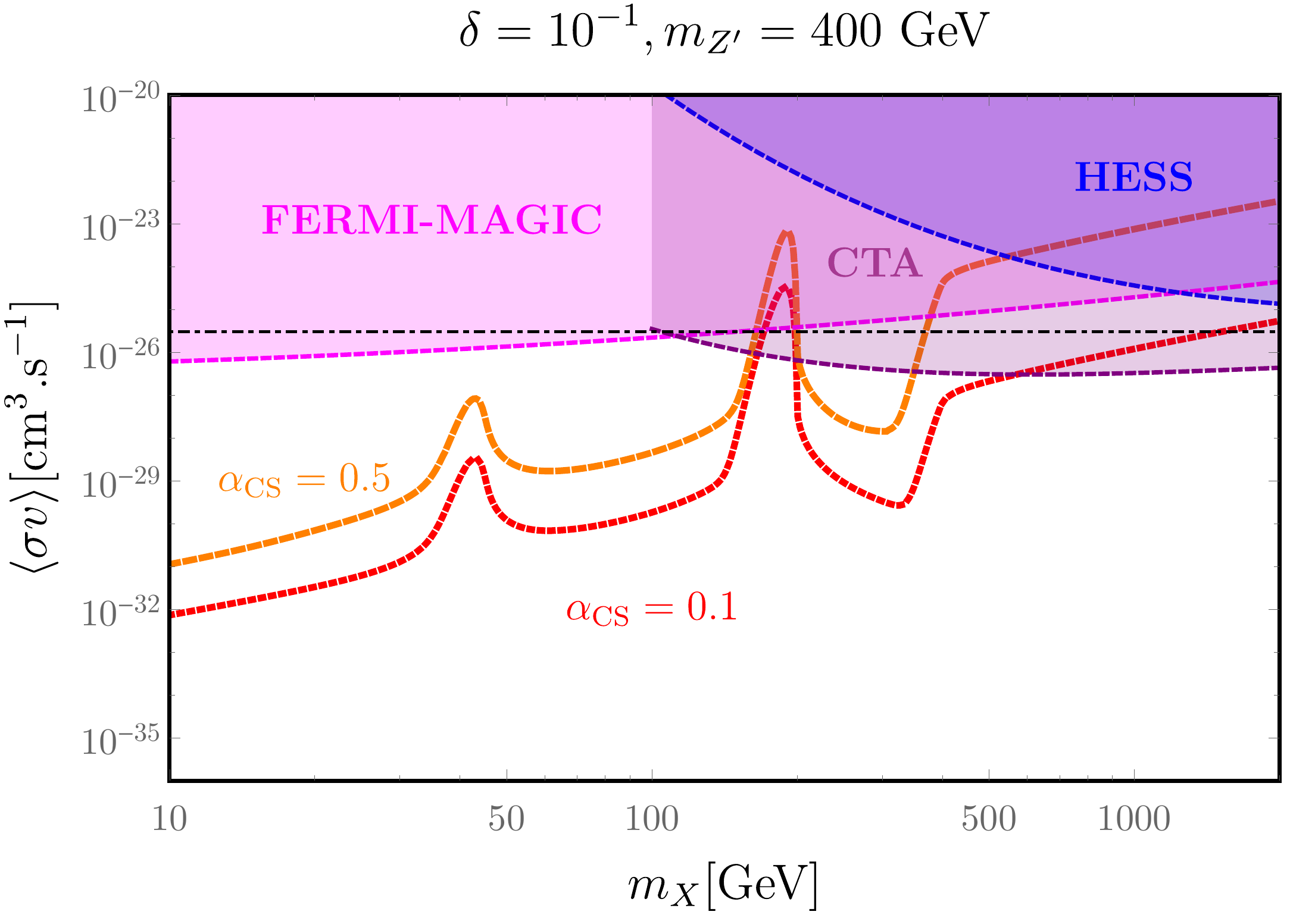}
\includegraphics[width=7.5cm]{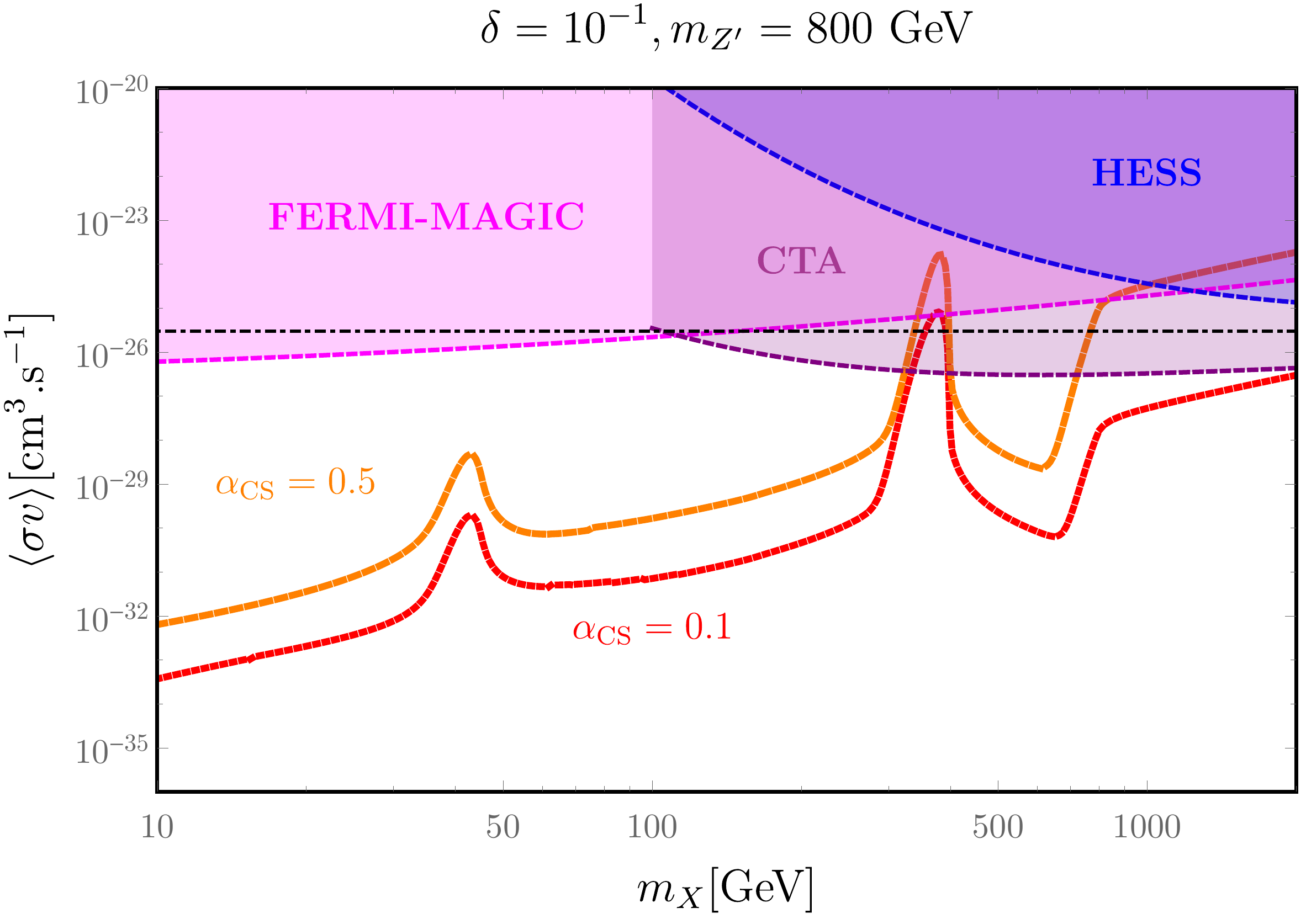}
\caption{Variation of the thermally averaged DM pair 
annihilation cross-section $\langle \sigma v\rangle$, at thermal freeze-out, 
as function of the DM mass $m_X$ for  
$m_{Z'}=400$ GeV (left) and $800$ GeV (right) with two values of $\alpha_{\rm CS}=0.1$
(red coloured solid line) and $0.5$ (orange coloured solid line)
keeping kinetic mixing parameter $\delta=0.1$. The black coloured dashed line represents the thermally 
favoured value of $\langle \sigma v\rangle=$  
$3 \times 10^{-26}\,{\mbox{cm}}^3\, {\mbox{s}}^{-1}$. The magenta 
coloured region with dashed outline represents the resultant exclusion from the combined FERMI-LAT and MAGIC observations 
(abbreviated as FERMI-MAGIC in all the successive relevant figures) while the blue coloured region
with dashed boundary represents the region already 
excluded by HESS for a Einasto density profile of the DM. Finally, the purple colour region
with dashed outline represents
the expected future exclusion from CTA.}
\label{fig:sigmav}
\end{figure}

The DM pair annihilation cross-section, as depicted in figure~\ref{fig:sigmav},
typically lies much below the thermally favoured value,
i.e., $10^{-{26}}\,{\rm cm}^3\,{\rm s}^{-1}$, ad exception of the pole regions 
$m_{X}\sim m_Z/2, m_{Z'}/2$, or DM masses above several hundreds of GeV so that at 
least annihilations into $ZZ'$ final state appears kinematically accessible.

\subsubsection{Indirect detection}

Residual annihilation processes of the DM at the present times could be 
efficient enough so that they might be detected by the Earth based telescopes or satellites. 
The absence of definite signals till date are translated into upper bounds 
on  $\langle \sigma v \rangle$ as a function of 
the DM mass. At the moment, the strongest limits come from searches of the gamma-rays produced in 
the DM annihilations. For DM masses below a few hundred GeVs, 
these limits are set by the FERMI 
satellite \cite{Ackermann:2015zua} and exclude the thermally favoured values of $\langle \sigma v\rangle$ for $m_X < 100$ 
GeV. For higher values of $m_X$ the best sensitivity is achieved by 
HESS \cite{Abramowski:2011hc} which has put a limit $\langle \sigma v \rangle \lesssim 
10^{-25}\text{cm}^3\,\text{s}^{-1}$ for $m_X \sim 1\,{\mbox{TeV}}$,
considering a Einasto density profile for the DM.

As shown earlier while discussing the relic density that most of the DM annihilation processes have a p-wave 
(or even d-wave) annihilation cross-section, i.e., they are velocity dependent. 
As a consequence the 
DM annihilation at present times is suppressed by several orders of magnitude with respect to 
its value at the thermal freeze-out and thus, limits from DM ID are actually not 
effective. ID can probe the WIMP paradigm for the DM relic density only 
when the latter is mostly determined by processes with an s-wave (i.e., velocity independent) 
annihilation cross-section. In our setup this requirement is fulfilled by annihilation into $Z'Z$ and $Z'Z'$ states (see eq.~(\ref{eq:XXzpz})
and eq.~(\ref{eq:XXzpzp})). The other s-wave 
dominated annihilation into $ZZ$ final state (see eq.~(\ref{eq:XXZZ})),
as already addressed, is insignificant
as it is suppressed by the fourth power of kinetic mixing parameter $\delta$.

We have thus compared the DM annihilation cross-section into $Z'Z$ and $Z'Z'$ 
final states with the limits derived from combined analysis of the MAGIC and FERMI-LAT observations
(abbreviated as FERMI-MAGIC) of the dwarf spheroidal galaxies (dSphs)~\cite{Ahnen:2016qkx}. We also consider limits from $10$ years of 
observations towards the inner galactic halo by HESS~\cite{Abdallah:2016ygi}. In the absence of a dedicated analysis for the 
considered final states we have applied the limit for gamma-rays 
originating from $W^+W^-$ pairs. This choice is reasonable since the $Z^\prime$ decays efficiently into hadrons as the SM 
gauge bosons. Any mild change in the limits as a result of this assumption is beyond the scope of our current work.

Our results are reported in figure~\ref{fig:sigmav} 
concerning changes of the DM pair annihilation cross-section at thermal freeze-out
with the DM mass $m_X$, for specific assignations of the other relevant inputs,
detailed in the figure caption. As evident from this figure that the impact of ID limits 
from FERMI-MAGIC (magenta coloured region) and HESS 
(blue coloured region) appears to be rather limited. This can 
be understood by looking at this simple analytical estimate for $\langle \sigma v \rangle_{Z'Z'}$:
%
\beq
\langle \sigma v\rangle_{Z^\prime Z^\prime}\simeq  \dfrac{8 \alpha_\text{CS}^4 
m_X^2}{9 \pi  m_{Z'}^4} \simeq 3\times 10^{-28} \text{cm}^3\,\text{s}^{-1} 
\Big( \dfrac{\alpha_\text{CS}}{0.1} \Big)^{4} \Big( \dfrac{m_X}{1 \text{ TeV}} 
\Big)^{2} \Big( \dfrac{m_{Z'}}{1 \text{ TeV}}\Big)^{-4},
\eeq
which shows that a value of the cross-section equal or bigger than the thermal expectation
i.e., $\sim\mathcal{O} (10^{-26}\, \text{cm}^3\,\text{s}^{-1})$,
can be achieved only for masses of the $Z'$ not exceeding a few hundreds of GeV and/or 
$\alpha_{\rm CS} \sim 0.5$. Such value of $\alpha_{\rm CS}$, however, is somewhat extreme since this coupling is expected 
to have a radiative origin as will be discussed later.
The impact of possible limits from ID will be anyway increased in the near 
future by Cherenkov Telescope Array a.k.a. CTA~\cite{Doro:2012xx,Pierre:2014tra,Wood:2013taa,Silverwood:2014yza,
Lefranc:2015pza,Lefranc:2016dgx} which we have reported
(purple coloured region) in figure~\ref{fig:sigmav},
assuming a projected limits from $500$~h of observation towards the galactic center.
  
\subsubsection{Direct detection}

DD experiments aim at measuring the recoil energy released by an
atomic nuclei upon a scattering process with a DM state. Several experiments have 
performed searches for DM scattering off nuclei reaching, in the last year, an impressive 
sensitivity. In particular, in the case of Spin Independent (SI) interactions of the 
DM with nuclei, a cross-section
$\sim\mathcal{O}(10^{-46}\text{ cm}^2)$, 
for DM mass of $50$ GeV, has been excluded by the LUX~\cite{Akerib:2016vxi} and 
PandaX~\cite{Tan:2016zwf}. These experiments play also a leading role in 
constraining the Spin Dependent (SD) interactions. In this case the maximal sensitivity is achieved for DM mass of 
$40$ GeV which corresponds to a value $\sim 10^{-41}\text{ cm}^2$ of the 
DM-nucleon scattering cross-section~\cite{Fu:2016ega}.

In the chosen setup (see eq.~(\ref{eq:starting_lagrangian})), scattering of \rm{the} DM with nucleons is originated, at the microscopic level, 
by an interaction between the DM and the SM quarks mediated via t-channel 
exchange of the $Z/Z'$ boson. An interaction of this kind, in the non-relativistic limit,
is described by the following effective operator~\cite{Belanger:2008sj}:

\begin{equation}
\mathcal{L}_{\rm DM~scattering}=\left(\frac{g_X^Z a_q^Z}{m_Z^2}
+\frac{g_X^{Z'} a_q^{Z'}}{m_{Z'}^2}\right) 
\left(\partial_\alpha X_\beta X_\nu-X_\beta \partial_\alpha X_\nu\right) 
\epsilon^{\alpha \beta \nu \mu}\ovl q \gamma_\mu \gamma_5 q,
\end{equation}
where $a_q^{Z'}=\dfrac{g^{Z^\prime}_{q_R}-g^{Z^\prime}_{q_L}}{2}$ 
and $a_q^Z=\dfrac{g^{Z}_{q_R}-g^{Z}_{q_L}}{2}$, which corresponds to a SD 
interaction with squared amplitude:
%
\beq
\overline{|\mathcal{M}|^2}=\dfrac{32 m^2_X m^2_N}{m^4_Z m^4_{Z^\prime}}
\Big( \sum_q a^{Z}_q \Delta_{q}^N g_{X}^{Z}m_{Z^\prime}^2+ 
\sum_q a^{Z'}_q \Delta_{q}^N g_{X}^{Z^\prime} m_{Z}^2 \Big)^2,
\eeq
with $m_N$ denoting the mass of a nucleon $N=p,\,n$ while $\Delta_q^N$ represents the contribution of the 
quark $q$ to the spin of the nucleon $N$. 
The SD scattering cross-section can be straightforwardly derived, taking into account the 
multiple isotopes present in the detector material as:
\beq
\sigma_{Xp}^\text{SD}=\dfrac{2}{\pi}\dfrac{\mu^2_{Xp}}{m_Z^4m_{Z^\prime}^4} 
\frac{ \sum\limits_A \eta_A \Big( S^A_{Z} g_{X}^{Z} m_{Z^\prime}^2+ 
S^A_{Z^\prime} g_{X}^{Z^\prime} m_{Z}^2 \Big)^2 }{\sum\limits_{A} \eta_A \Big( S^A_n + S^A_p \Big)^2},
\eeq
where $\mu_{Xp}=m_X m_p/(m_X+m_p)$ is the reduced mass for 
the DM-proton system and $S^A_{Z^{(\prime)}}=a^{Z^{(\prime)}}_u(\Delta^p_u S^A_p+\Delta^p_d S^A_n)+
a^{Z^{(\prime)}}_d[(\Delta^p_d+\Delta^p_s)S^A_p+(\Delta^p_u+\Delta^p_s)S^A_n]$. 
Here $S^A_{p,n}$ represents the contribution of protons and neutrons to the spin of 
a nucleus with atomic number $A$ while $\eta_A$ represents the relative abundance 
of a given isotope of the element constituting the target material. Notice that the 
result is almost independent of the DM mass since the only dependence 
through $\mu_{Xp}$ will vanish for $m_X \gg m_p$ giving
$\mu_{Xp} \sim m_p$. A simple estimate 
of $\sigma_{Xp}^\text{SD}$ can be 
performed assuming $\delta,\alpha_\text{CS} \ll 1$ as:
\beq
\sigma_{Xp}^\text{SD}\simeq 6 \times 10^{-50} \text{cm}^2 \Big( \dfrac{\delta}{0.1} 
\Big)^2 \Big( \dfrac{\alpha_\text{CS}}{0.1}  \Big)^2 
\Big( \dfrac{m_{Z^\prime}}{1 \text{ TeV}}  \Big)^{-4},
\eeq
which gives values of $\sigma_{Xp}^\text{SD}$ well below the 
current maximal sensitivity ($10^{-41} \text{cm}^2$),  and remains also beyond the reach of next generation detectors
for the chosen set of parameter values, consistent with 
other existing constraints.

\subsection{Collider phenomenology}

Phenomenological models, where
the DM candidate interacts with the SM fields through a spin-1 mediator 
typically posses a rich collider phenomenology. In the case when 
on-shell production of a $Z'$ is kinematically accessible
in proton-proton collision, detectable signals can appear from the 
decays of $Z'$ into SM fermions, showing a new resonance 
in the invariant mass distribution of dijets~\cite{Khachatryan:2015dcf,ATLAS:2015nsi,Sirunyan:2016iap,Aaboud:2017yvp} or 
dileptons~\cite{Aaboud:2016cth,Khachatryan:2016zqb,ATLAS:2017wce} as well as from possible 
decay of a $Z'$ into DM pairs which can be probed through mono-{\bf{X}} searches 
({\bf{X}}= {hadronic jets}, photon, weak gauge bosons, SM-Higgs)~\cite{
Sirunyan:2017hci,Aaboud:2016qgg,CMS:2016hmx,Aaboud:2016uro,CMS:2016fnh,Aaboud:2017dor,
ATLAS:2017uwx,ATLAS:2017pqx,Sirunyan:2017hnk} accompanied by moderate/large missing transverse
energy/momentum. Interestingly, the relative relevance of these two 
kinds of searches, i.e., resonances and mono-{\bf X} events, is mainly set by the invisible decay branching 
fraction (Br) of the $Z'$ which, in turn, is constrained by the DM observables 
that primarily appears from the requirement of correct relic density for the chosen framework.

\begin{figure}[t!]
\centering
\includegraphics[width=7.5 cm]{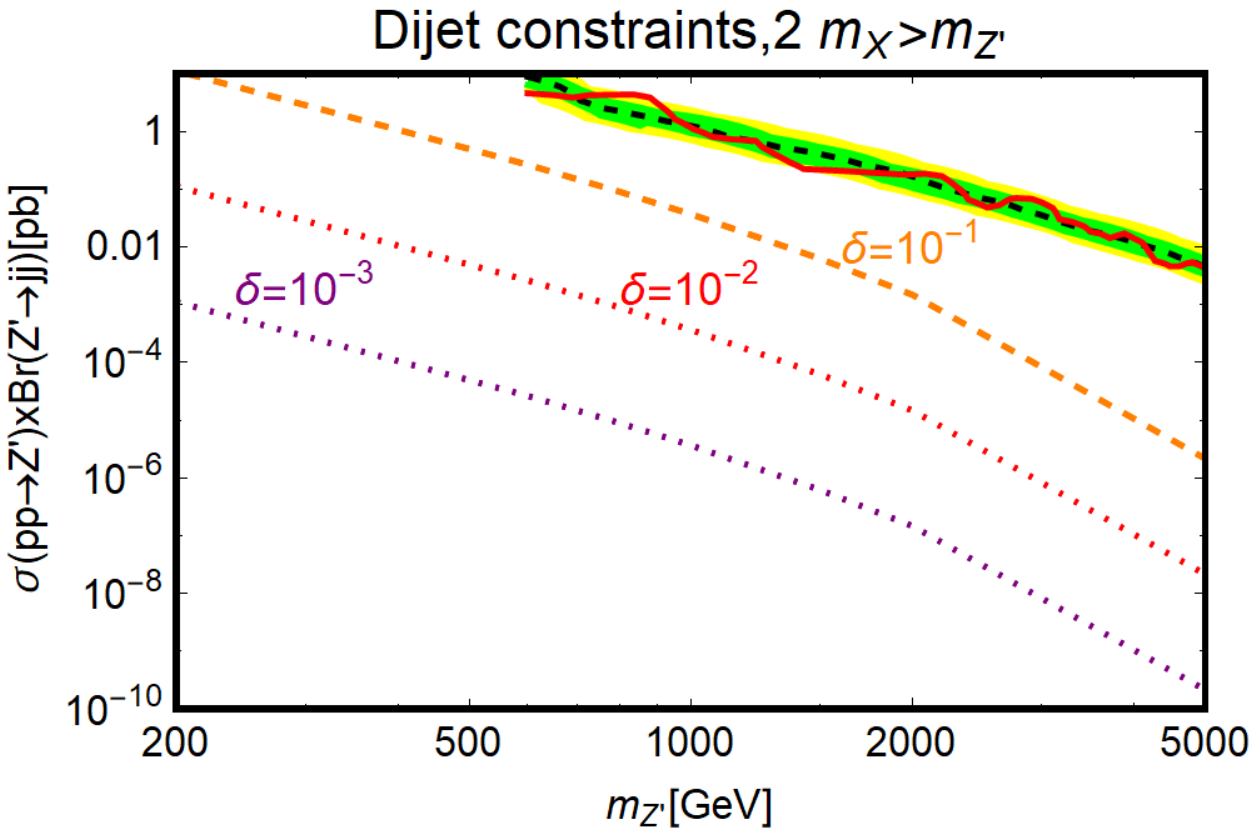}
\includegraphics[width=7.5 cm]{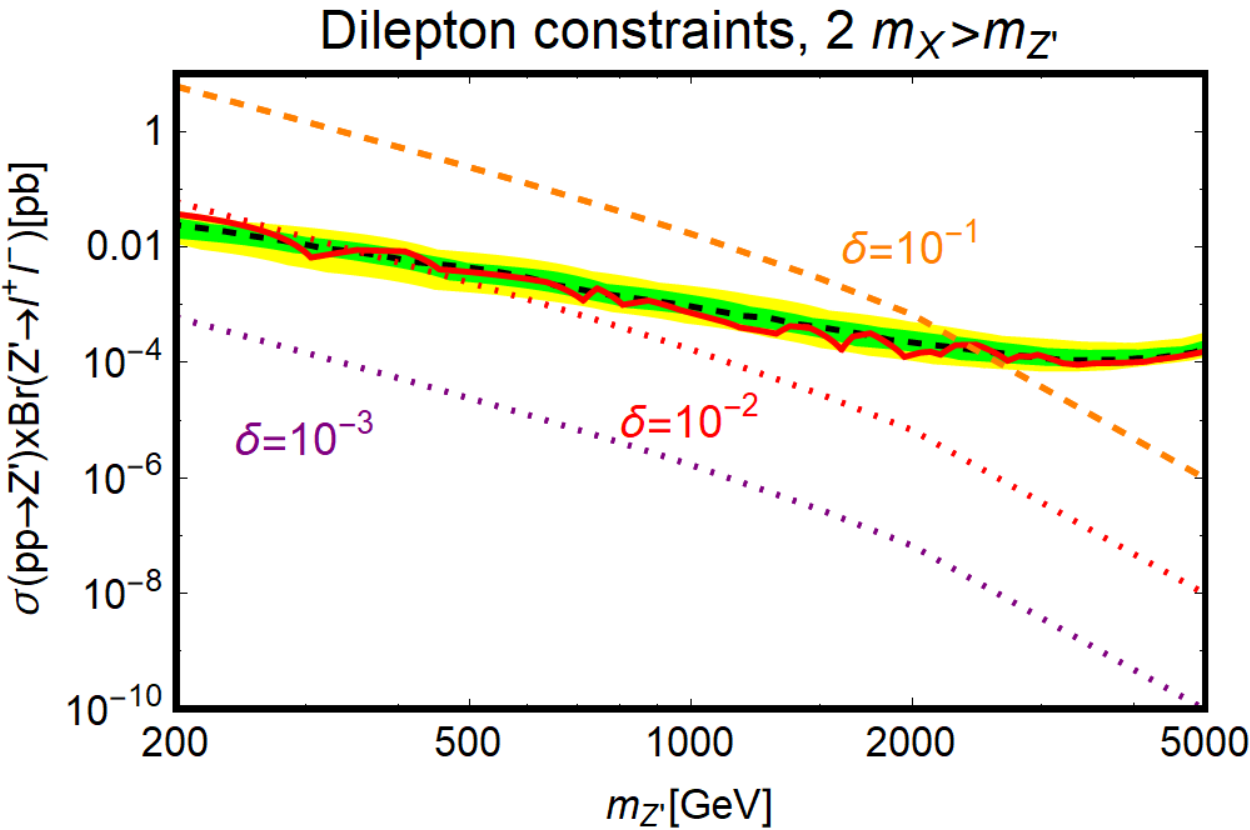}
\caption{Comparison of the production cross-sections for 
$pp\to Z' \to jj$ (left panel) and  $pp\to Z' \to l^+ l^-$ (right panel)
processes for an on-shell $Z'$ in the chosen theory framework
with the relevant experimental limits~\cite{Khachatryan:2015dcf} 
and~\cite{ATLAS:2017wce}, respectively, as a function
of $m_{Z'}$ for three fixed
values of $\delta=10^{-3},\,10^{-2}$ and $10^{-1}$.
Here the solid red coloured
curve represents the observed experimental limit while 
dashed black coloured line, green and yellow coloured bands represent
the expected limit and its $1\sigma,\,2\sigma$ ranges. The dotted
purple, red and dashed orange coloured lines are used for 
$\delta=0.001,\,0.01$ and $0.1$ configurations, respectively.
For simplicity, we have considered 
the case of $2 m_X > m_{Z'}$ so that $Z'$ has only visible decays.
In our analysis we consider $j=u,\,d,\,c,\,s,\,b$ and 
$t$ for $m_{Z'}> 2m_t$ and $l=e,\,\mu$.}
\label{fig:collider1}
\end{figure}

The impact of experimental limits from resonance searches is
depicted in figure~\ref{fig:collider1} where, for simplicity, we consider
$2m_X > m_{Z'}$ to forbid $Z'\to$ DM pairs process.
For these analyses we consider all possible quark flavours
including top for $m_{Z'}> 2m_t$ regime while $l=e$ and $\mu$ only.
Here we have compared the results of numerical
simulations for the chosen setup with the experimental ones
including $1\sigma,\,2\sigma$ variations of the production
cross-sections as observed in refs.~\cite{Khachatryan:2015dcf} 
and~\cite{ATLAS:2017wce}, respectively.
Since the $Z'$-SM fermions mixing (see eq.~(\ref{eq:zzpsmfermion})) appears from an effective
$Z$-$Z'$ mixing, both $\sigma(pp\to Z')$ and Br($Z\to jj/l^+l^-$) are 
sensitive to the parameter $\delta$. The resultant $\delta^2$ dependence
thus, hints diminishing $\sigma(pp\to Z')\times {\rm Br} (Z'\to jj/l^+l^-)$ values
for decreasing $\delta$ values,
as also reflected in figure \ref{fig:collider1}. It is evident from figure \ref{fig:collider1} that
once values of the kinetic mixing parameter $\delta$ smaller 
than $\mathcal{O}(1)$ are considered to comply with the theoretical and EWPT 
(see eq.~(\ref{eq:EWPT})) constraints, only the limits 
from dileptons resonance searches remain effective. 
For $\delta=0.1$, values of $m_{Z'}$ below 2 TeV 
are ruled out while for $\delta=0.01$, a much weaker lower bound of approximately 300 GeV is obtained
on $m_{Z'}$. For further lower values of $\delta$, very small $\delta^2$
dependence\footnote{Holds true for a resonant production with no/suppressed invisible decay of $Z'$.} makes 
$m_{Z'}$ unconstrained from the aforesaid searches.

In the presence of a non-zero and sizable invisible branching fraction for $Z'$, i.e., Br($Z'\to$ DM pairs),
the production cross-sections of dijets/dileptons get suppressed by 
the enhanced decay width of $Z'$. In such a scenario, the 
cross-section corresponding to mono-{\bf X} signals might become sizable, possibly providing complementary 
constraints. We preserve the discussion of such complementary signals for the next subsection.

\subsection{Results}

\begin{figure}[t!]
\begin{center}
\includegraphics[width=8 cm]{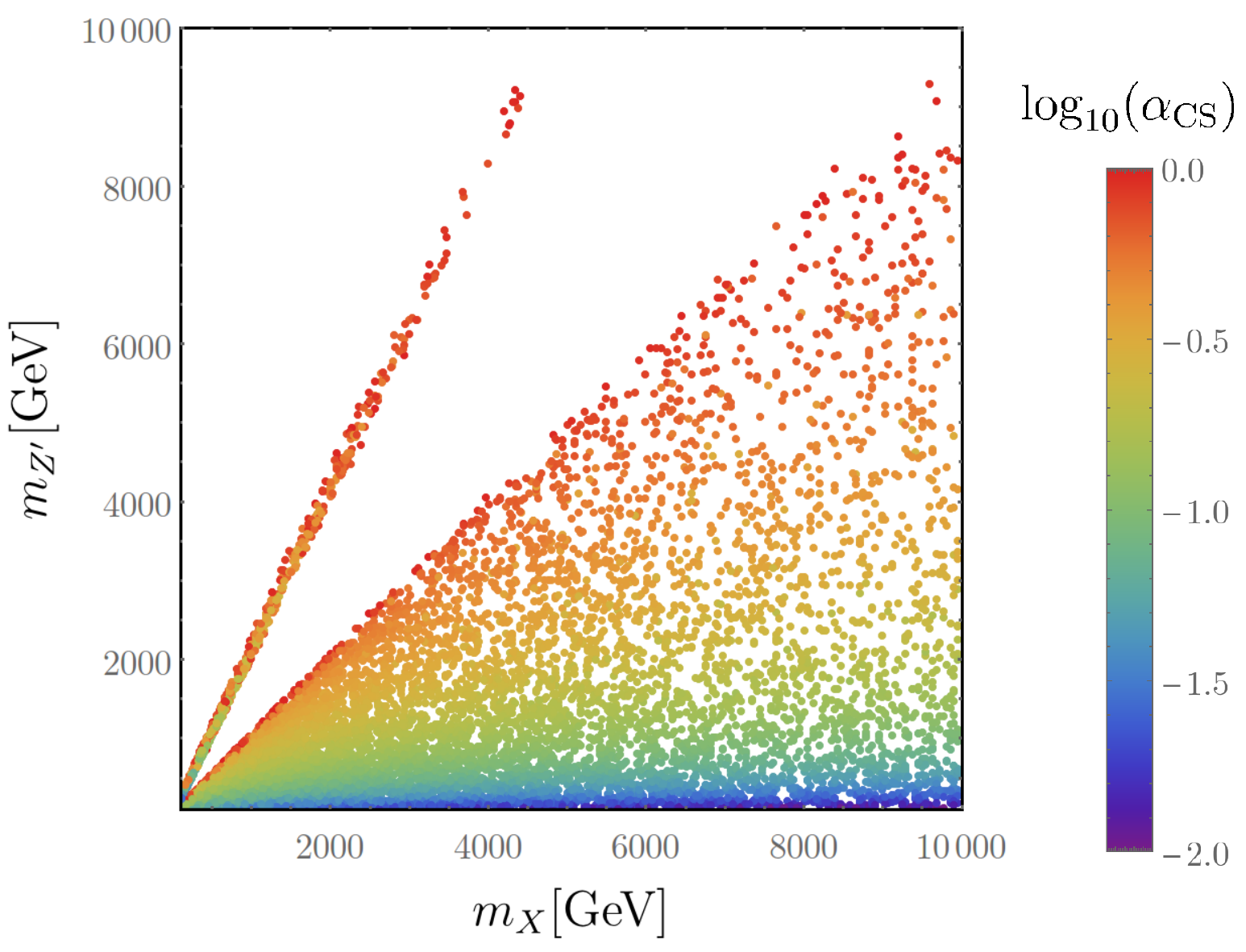}
\caption{Summary of results in the bi-dimensional plane $m_X,m_{Z'}$
for a vectorial DM with $Z'$-portal in the presence
of CS interaction and kinetic mixing terms.
The colour code corresponds to the variation in $\alpha_{\rm CS}$ values.
The plot reports model points, generated through a parameter scan, 
illustrated in the main text, passing constraints from the DM phenomenology 
and general constraints on the kinetic mixing parameter.}
\label{fig:scan}
\end{center}
\end{figure}

In this subsection we report the impact of various constraints, as mentioned in the previous subsections, on the parameter space of the model.
As already pointed out the effect of DD and ID constraints on our
model framework is substantially negligible. Concerning the DM phenomenology, 
the only relevant constraint comes from the requirement of the correct relic density. We 
have then conducted a more extensive analysis, with respect to the one presented in figure~\ref{fig:sigmav} 
by performing a scan over the four free input parameters in the following ranges:
\beq
\label{eq:scan1}
\delta \in \left[10^{-3}, 1\right], \,\,\,\,
\alpha_{\rm CS} \in \left[10^{-2}, 1\right],\,\,\,\,
m_X \in \left[100\,\mbox{GeV},10\,\mbox{TeV}\right],\,\,\,\,
m_{Z^\prime} \in \left[100\,\mbox{GeV},10\,\mbox{TeV}\right],
\eeq
and retaining the model points featuring the correct DM relic density\footnote{We consider points 
corresponding to $\Omega_X h^2 = 0.12 \pm 10\%$ variation.} and respecting, at the same time, 
constraints from the EWPT, SM $\rho$-parameter measurement as well as reproducing experimentally viable mass 
and width for the $Z$-boson.
The ensemble of points respecting these aforementioned constraints has been reported in figure~\ref{fig:scan}, 
in the $m_X$-$m_{Z'}$ bi-dimensional plane with a colour code showing variations in $\alpha_{\rm CS}$ values.


\begin{figure}[t!]
\centering
\includegraphics[width=5.0 cm]{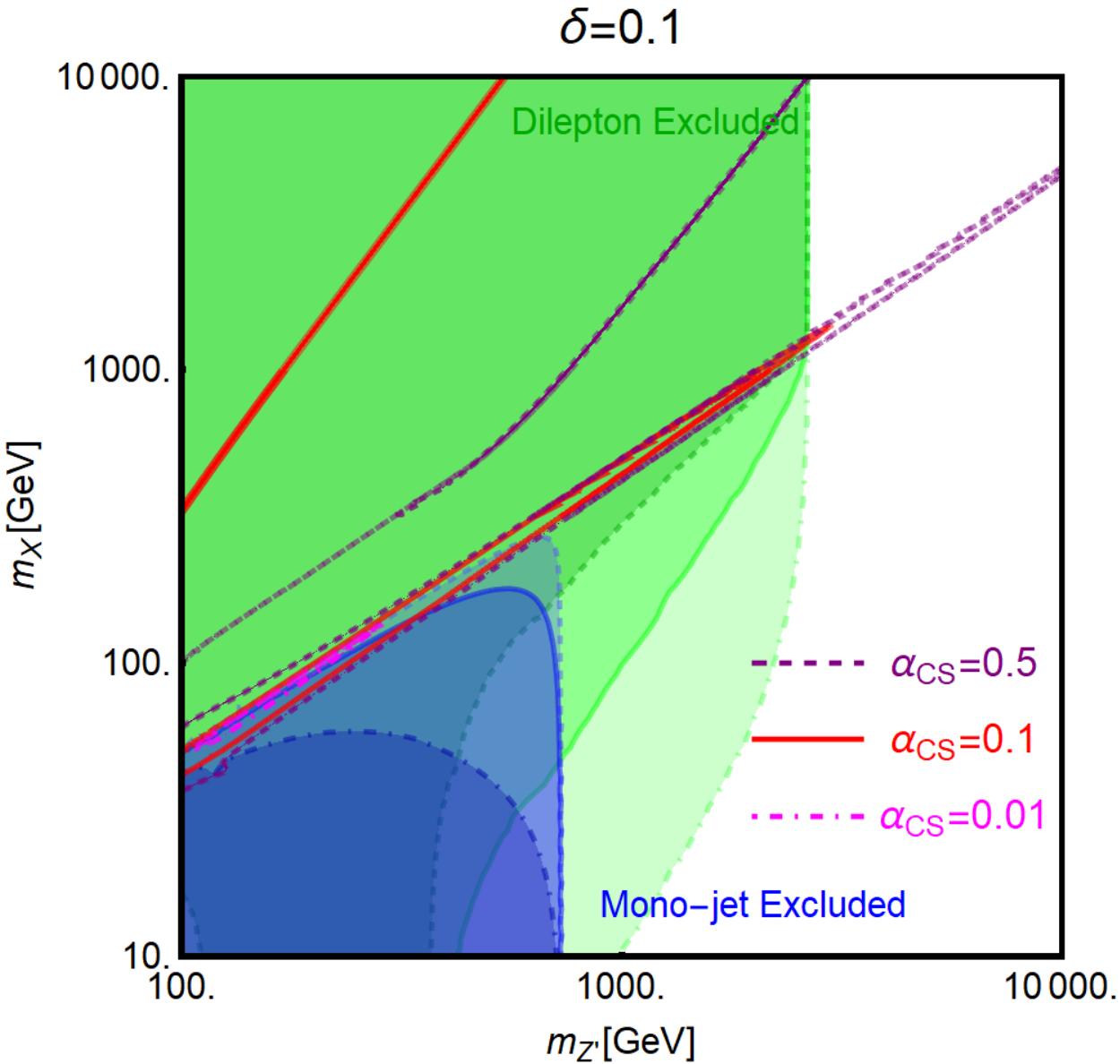}
\includegraphics[width=5.0 cm]{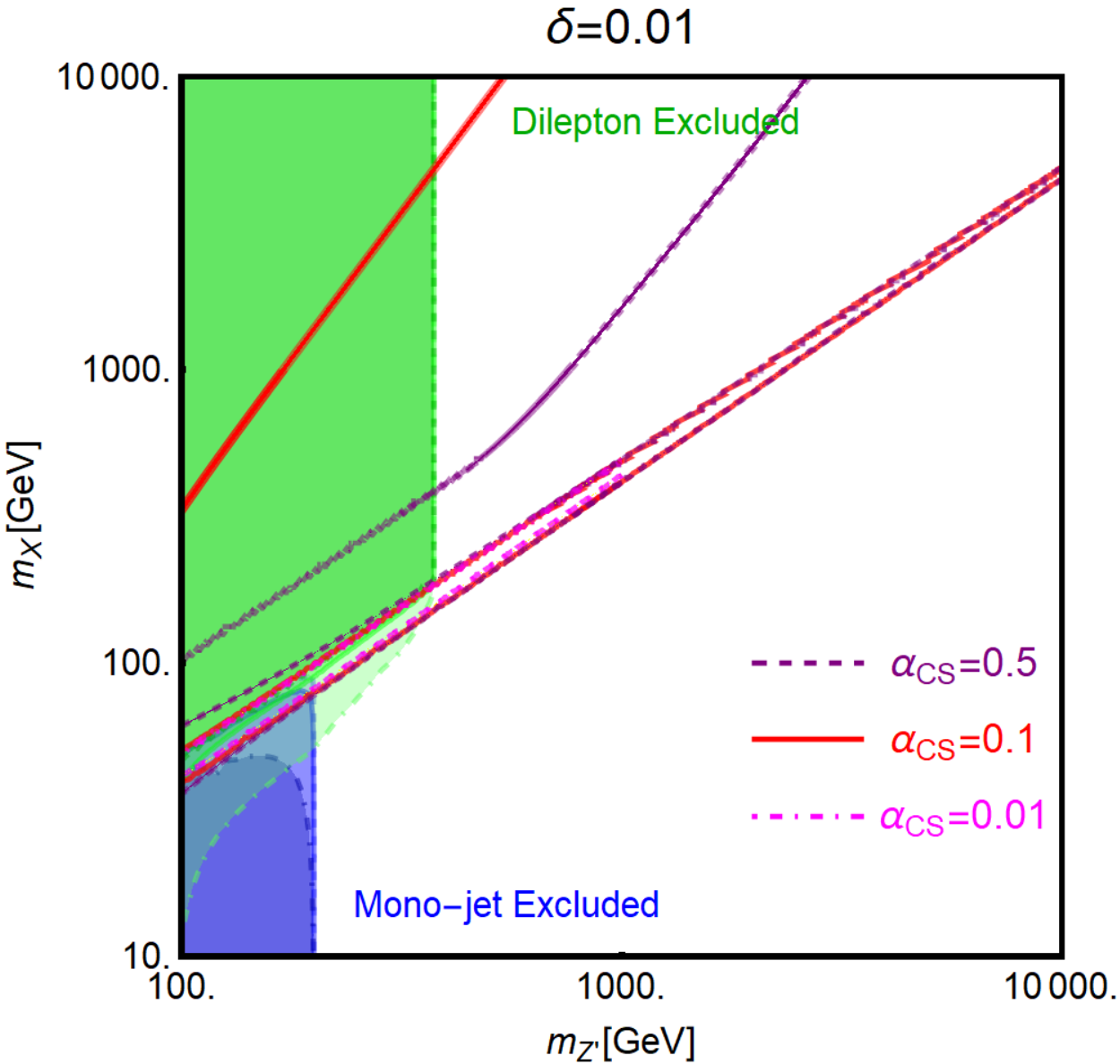}
\includegraphics[width=5.0 cm]{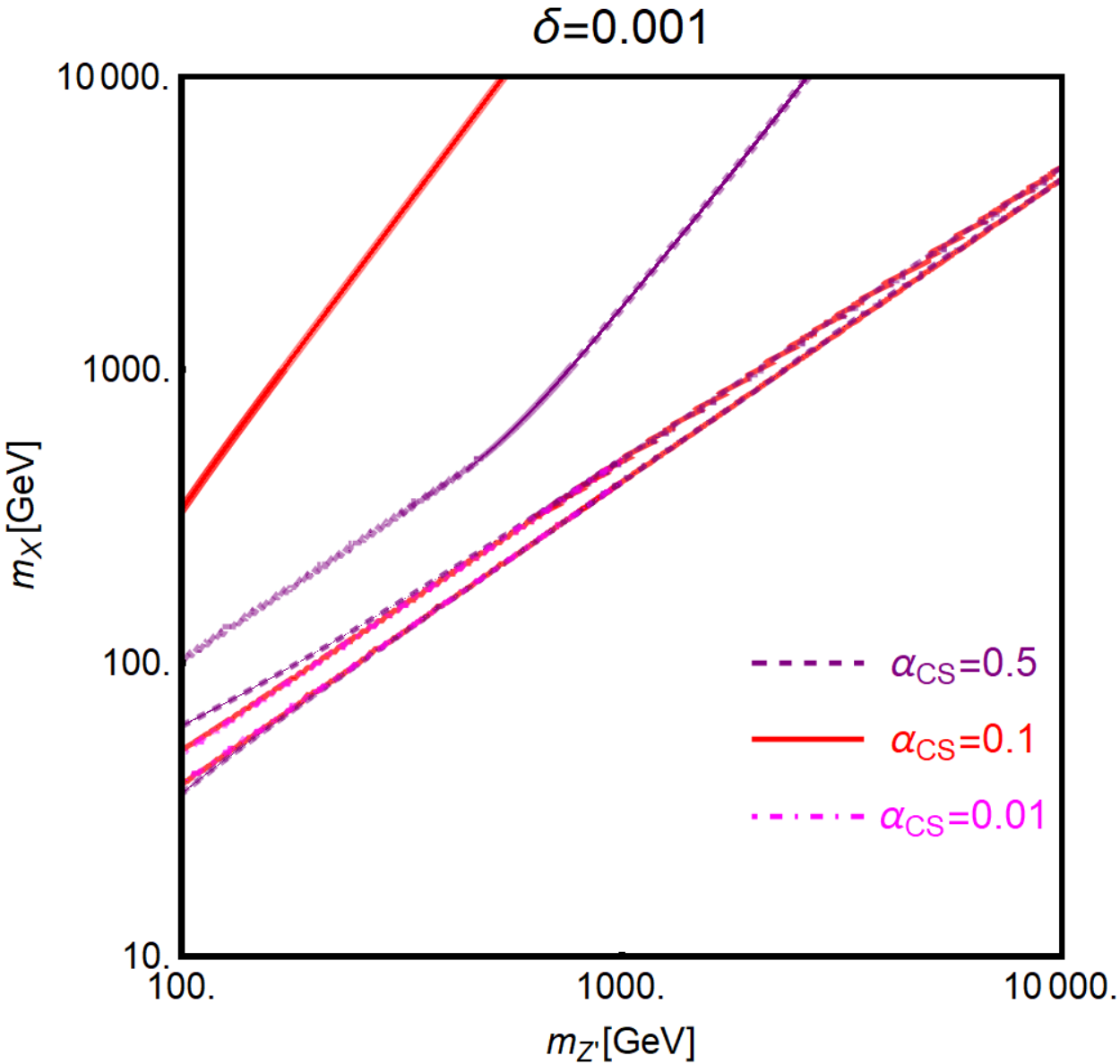}
\caption{Summary of the relic density and the collider constraints in the $(m_{Z'},m_X)$ bi-dimensional plane 
for the three values of $\delta$, $0.1$ (left), $0.01$ (middle) and $0.001$ (right) choosing
three assignations of $\alpha_{\rm CS}$, namely $0.5$, $0.1$, and $0.01$. 
In each plots the dashed magenta, solid red and dot-dashed purple coloured curves 
represent the correct DM relic density for $\alpha_{\rm CS}=0.5,\,0.1$ and $0.01$, respectively. 
The regions covered by the light blue colour with dashed boundary, blue colour with solid outline 
and dark blue colour with dot-dashed boundary
are excluded by mono-jet searches when one considers
the value of $\alpha_{\rm CS}=0.5,\,0.1$ and $0.01$, respectively.
The same set of $\alpha_{\rm CS}$ values is associated with 
the three different regions, namely, 
dark green coloured with dashed boundary, green coloured
with solid outline and light green coloured with dot-dashed boundary, respectively, that are excluded from the 
searches of dilepton resonances.}
\label{fig:collider_summary}
\end{figure}

%
As evident from figure \ref{fig:scan}, in agreement with the general discussion of 
sub-subsection \ref{sssec:rdsc1}, that the correct relic density can be achieved 
either in the  ``pole'' region $m_X \sim \frac{m_{Z'}}{2}$ or for $m_X \gtrsim m_{Z'}$, when the process 
$XX \rightarrow Z'Z'$ is kinematically allowed.\footnote{As can be noticed from the analytical expression 
(see eq.~(\ref{eq:XXzpzp})) that the annihilation cross-section into $Z'Z'$ increases with the DM mass. 
This is somewhat an unreasonable behavior possibly leading to the violation of unitarity (see, for example, 
refs.~\cite{Griest:1989wd,Beacom:2006tt,Englert:2016joy,Kahlhoefer:2015bea,Arcadi:2017atc}). We emphasize 
here that the studied phenomenological Lagrangian is actually the low-energy effective theory limit of a more 
complete framework. Thus, violation of unitarity for some assignations of the model parameters would simply imply 
that for those parameter values the effective field theory limit is invalid and hence, additional degrees of 
freedom should be included to develop a UV complete theory. We have verified that possible consequence of the 
unitarity violation would affect the thermal DM region only when $\alpha_{\rm CS}\gtrsim \mathcal{O}(1)$. The 
results of our findings as reported in figure~\ref{fig:collider_summary} are instead not affected by this limit 
as we have considered lower values of $\alpha_{\rm CS}$ parameter.} 
%
We have successively examined the impact of collider constraints on our construction 
in figure~\ref{fig:collider_summary}. For a better and elucidate illustration
of our findings, we have considered three fixed assignations of $\delta$, namely $0.001,\,0.01$ and 
$0.1$ and the same for $\alpha_{\rm CS}$, $0.01,\,0.1$ and $0.5$ keeping $m_X$ and $m_{Z'}$ as 
the two free varying input parameters. The relevant results are
reported in the three panels of figure~\ref{fig:collider_summary} corresponding to the three different 
values of $\delta$, namely $0.1$ (left), $0.01$ (middle)
and $0.001$ (right) where isocontours of the correct DM relic density 
are shown for the three assignations of $\alpha_{\rm CS}$ parameter using three different
representations (purple coloured dot-dashed line for $\alpha_{\rm CS}=0.01$, 
solid red coloured line for $\alpha_{\rm CS}=0.1$ and magenta coloured dashed line for $\alpha_{\rm CS}=0.5$).
These isocontours have been compared with the limits from  
dilepton resonances and mono-jet searches. The 
former is evaluated in a similar fashion 
as of figure~\ref{fig:collider1} by computing the associated cross-section, 
as a function of the input parameters $\delta,\alpha_{\rm CS}, m_X$ and $m_{Z'}$ using the package 
MadGraph5$\_$aMC$@$NLO~\cite{Alwall:2014hca} and compared with the relevant
experimental observations. Contrary 
to what we did earlier, now we have also accounted 
for the possibility of a sizable invisible branching fraction 
of $Z'$ by suitably rescaling the experimental limit according to the procedure illustrated in
ref.~\cite{Arcadi:2013qia}. 
As evidenced from the left panel of figure~\ref{fig:collider_summary} 
that the previously quoted limit of approximately 2 TeV (see figure \ref{fig:collider1}, right panel plot) 
for $\delta=0.1$ is actually effective only when $m_{Z'}< 2 m_X$. When  
decay of a $Z'$ into DM pairs is kinematically accessible, i.e., $m_{Z'}> 2 m_X$,
and $\alpha_{\rm CS} \gtrsim \mathcal{O}(1)$, the lower limit on $m_{Z'}$ 
can be reduced even to a few hundreds 
of GeV. The mono-jet limits have been derived by evaluating the production cross-section times the detector efficiency 
and acceptance in the selection of the final state by generating and analyzing events, corresponding to the process 
$p p \rightarrow XXj$, through the combination of MadGraph5$\_$aMC$@$NLO~\cite{Alwall:2014hca} 
(matrix element calculation) PYTHIA 
8~\cite{Sjostrand:2014zea} (event generation and hadronization) and 
DELPHES 3 (fast detector simulations)~\cite{deFavereau:2013fsa}. 
The selection acceptance for the generated events has been determined by 
imposing a minimal value of 500 GeV 
for the missing transverse energy. 
The mono-jet limits have been determined by imposing 
an upper bound of $\sim 6$ fb on the product of production cross-section,
signal acceptance and the detection efficiency \cite{Sirunyan:2016iap}
Our procedure has been validated by reproducing the 
excluded region for the benchmark model adopted in ref.~\cite{Sirunyan:2016iap}.

It is apparent from figure \ref{fig:collider_summary} that
both kinds of the collider limits, i.e., dileptons and mono-jet, are essentially effective for large and
moderate values of $\delta$, i.e., significant
for $\delta=0.1$ (left panel) and moderate for 
$\delta=0.01$ (middle panel).
These behaviours, as already stated,
are expected since the parameter $\delta$ determines the production vertex of a 
$Z'$ as well as its decay rate into the SM fermions. Hence, 
collider production cross-section suppresses very fast 
as the kinetic mixing parameter decreases. On the 
contrary, the requirement of the correct relic density is more moderately affected by the decrease 
in the value of the kinetic 
mixing parameter. This happens because the correct relic density 
is achieved mostly through the annihilation into $Z'Z'$ final states, whose rate does 
not depend on $\delta$, or in the ``pole'' region, where variations of the couplings can be compensated by slight 
changes of $|2m_X-m_{Z'}|$. 

Among the different collider constraints the most effective ones are the ones 
which emerge from searches of dilepton resonances, 
even if the invisible decay channel for a $Z'$ is taken into account. Using the invariant mass distribution of 
the heavy dilepton resonance, peaked at the $Z'$ mass, one can discriminate the signal 
from the background nicely and for this reason the dilepton channel is 
considered to be a very good probe for these kinds of models.

As evidenced from the left and middle plots of 
figure~\ref{fig:collider_summary} that these searches
exclude most of the viable thermal DM region, leaving just a small portion 
of the parameter space around the $Z'$ pole region. On the contrary, the impact of the mono-jet 
constraints is much more moderate and exists only 
for the light DM masses. The collider constraints, as already discussed, 
disappear very fast as the 
value of the kinetic mixing parameter $\delta$ decreases. For $\delta=0.01$,
(middle panel of figure \ref{fig:collider_summary}),
compared to the $\delta=0.1$ scenario,
a much smaller portion of the parameter space
remains excluded from the dileptons and mono-jet
constraints.
All the collider constraints disappear for $\delta=0.001$ 
(right panel of figure \ref{fig:collider_summary}) where the only constraint on 
the model parameter space comes from the requirement of the correct DM relic density.

%
\section{Scenario-II: $Z'$-$Z$ interaction via a second Chern-Simons term}
\label{sec:scenario2}

In this section, just like section \ref{sec:scenario1},
we consider a CS term to connect a vectorial DM
with a $Z'$. However, unlike  section \ref{sec:scenario1}
the coupling between the $Z'$ mediator and the neutral 
EW gauge bosons arises via a second independent CS term.
Being guided by our previous approach, we will address the phenomenological implications of this 
setup as the result of a numerical scan over the free inputs after a brief illustration of the model followed
by concise discussions on the constraints of correct relic density
and different DM searches.

\subsection{The Lagrangian}

The Lagrangian describing the low-energy phenomenology of the 
aforementioned setup can be written as:
%
\begin{equation}
\mathcal{L}\supset \alpha_{\rm CS} \epsilon^{\mu \nu \rho \sigma}  X_{\mu} Z^\prime_{\nu} 
X_{\rho \sigma} + \beta_\text{CS} \epsilon^{\mu \nu \rho \sigma}  
Z_{\mu} Z^\prime_{\nu} B_{\rho \sigma}  +\frac{m_{Z^\prime}^2}{2}Z^{\prime \mu} 
Z^\prime_\mu+\frac{m_{X}^2}{2}X^\mu X_\mu,
\label{eq:lagdoubleCS}
\end{equation}
where $\alpha_\text{CS}$ and $\beta_\text{CS}$ are the $XXZ^\prime$ and $ZZZ',\,Z\gamma Z'$ coupling constants, 
respectively.  $B_{\rho \sigma}$ denotes the field strength of the SM hypercharge gauge field. The origin 
of the coupling $\alpha_\text{CS}$ is the same as of section \ref{sec:scenario1} which will be discussed 
later in appendix \ref{sec:appendixC}. The second CS coupling $\beta_\text{CS}$ is non-invariant under the SM gauge 
group transformation. Nevertheless, as pointed out in ref.~\cite{Dudas:2012pb}, it could be generated by considering 
the following gauge invariant effective operator obtained after integrating out some heavy degrees of freedom:

\begin{equation}
\label{eq:twoCSLorigin}
\mathcal{L}\propto i\, \epsilon^{\mu \nu \rho \sigma} D_\mu 
\theta_{Z^\prime}\Big( (D_\nu H)^\dagger H - H^\dagger D_\nu H \Big)B_{\rho \sigma},
\end{equation}
where $D_\mu\theta_{Z^\prime}=\partial_\mu \theta_{Z^\prime}-v_{Z^\prime}q_{Z^\prime}g_{Z^\prime} Z^\prime_\mu$ \
represents the covariant derivative of Stueckelberg axion $\theta_{Z^\prime}$ while $D_\nu H$ denotes the usual 
covariant derivative of the SM-Higgs doublet. $q_{Z^\prime},\,v_{Z^\prime}$ are the charge and VEV of the associated
complex scalar field and $Z'_\nu$ is the gauge boson of the concerned
$U(1)_{Z^\prime}$ group with $g_{Z^\prime}$ as the gauge coupling.
After the EWSB and choosing unitary gauge for the $U(1)_{Z'}$ group
(such that $\theta_{Z^\prime}$, connected to the phase of an associated heavy Higgs 
field, gets ``eaten" by the longitudinal component of $Z'_\mu$), we recover the operator considered in eq.~(\ref{eq:lagdoubleCS}).

The structure of eq.~(\ref{eq:lagdoubleCS})
contains two CS couplings, namely $\alpha_{\rm CS}$ and $\beta_{\rm CS}$. The former is the same as of 
eq.~(\ref{eq:starting_lagrangian}) whose origin is explained in the appendix \ref{sec:appendixC} while the latter
appears from eq.~(\ref{eq:twoCSLorigin}). Given that
the relevant charges are $\sim\mathcal{O}(1)$ (see eq.~(\ref{eq:effctivealphaCS}), the parameter
$\alpha_{\rm CS}$ from the associated pre-factor goes
as $\sim\mathcal{O}(10^{-3})$. On the other hand, from eq.~(\ref{eq:twoCSLorigin}) the pre-factor for $\beta_{\rm CS}$ 
varies as $v_{Z^\prime} v_h^2/M^3$ 
with $M$ as the cut-off scale of the theory,
related to the mass of the associated heavy BSM fermions. Now assuming other relevant parameters as $\sim \mathcal{O}(1)$, 
even when $v_{Z^\prime}\sim M$, one gets
$\beta_{\rm CS}\sim v_h^2/M^2$. Hence, for $v_h \sim \mathcal{O}(10^2$ GeV) and $M\sim \mathcal{O}(10$ TeV)
(this conservative limit is consistent with
the hitherto undetected evidence of BSM physics at the
13 TeV LHC operation), $\beta_{\rm CS}\sim \mathcal{O}(10^{-4})$ or $0.1 \times \alpha_{\rm CS}$. It is
thus apparent that in general a hierarchy between the values of two CS couplings is rather natural as they have 
two different theory origins. In our analysis
we, however, also consider the possibility of $\alpha_{\rm CS}=\beta_{\rm CS}$. 

\subsection{Relic density}

In the case of double CS terms, unlike the kinetic
mixing scenario, the number of accessible DM pair annihilation channels 
is limited just to three options, i.e., $Z\gamma$, $ZZ$ and $Z'Z'$. The first
two are induced by s-channel exchange of the $Z'$ while the third one is 
induced by t/u channel exchange of a DM state. Simple analytical approximations of 
the corresponding DM pair annihilation cross-sections can be
derived, as usual, through the customary velocity expansion:

\begin{itemize}
\item $XX\to Z\gamma:$
\beq
\label{eq:XXZgam}
\langle \sigma v\rangle_{Z\gamma}\simeq \frac{\alpha_\text{CS} ^2 
\beta_\text{CS} ^2 c_W^2 \left(4 m_X^2-m_Z^2\right)^3}{48 \pi  m_X^4 m_{Z^\prime}^4}\dfrac{1}{x},
~~{\rm for~} m_Z \ll m_X .
\eeq
%
\item $XX\to ZZ:$
\beq
\label{eq:XXZZ2}
\langle \sigma v\rangle_{ZZ}\simeq \frac{8 \alpha_\text{CS} ^2 
\beta_\text{CS} ^2 s_W^2 \left(m_X^2-m_Z^2\right)^{3/2} }{3 \pi m_X  m_{Z^\prime}^4}\dfrac{1}{x},
~~{\rm for~} m_Z \ll m_X .
\eeq
%
\item $XX\to Z^\prime Z^\prime:$
\bea
\label{eq:XXzpzp2}
\langle \sigma v\rangle_{Z^\prime Z^\prime}&&\simeq \frac{\alpha_\text{CS} ^4  
\left(32 m_X^8+14 m_{Z^\prime}^8-56 m_X^6 m_{Z^\prime}^2+69 m_X^4 m_{Z^\prime}^4-50 m_X^2 m_{Z^\prime}^6\right)}{9 \pi  m_X^2 m_{Z^\prime}^4 \left(m_{Z^\prime}^2-2 m_X^2\right)^2} 
\sqrt{1-\frac{m_{Z^\prime}^2}{m_X^2}}, \nonumber\\
&&~~~~{\rm for~} m_{Z^\prime} \ll m_X .
\eea
\end{itemize}

\begin{figure}[t]
\centering
\includegraphics[width=8 cm]{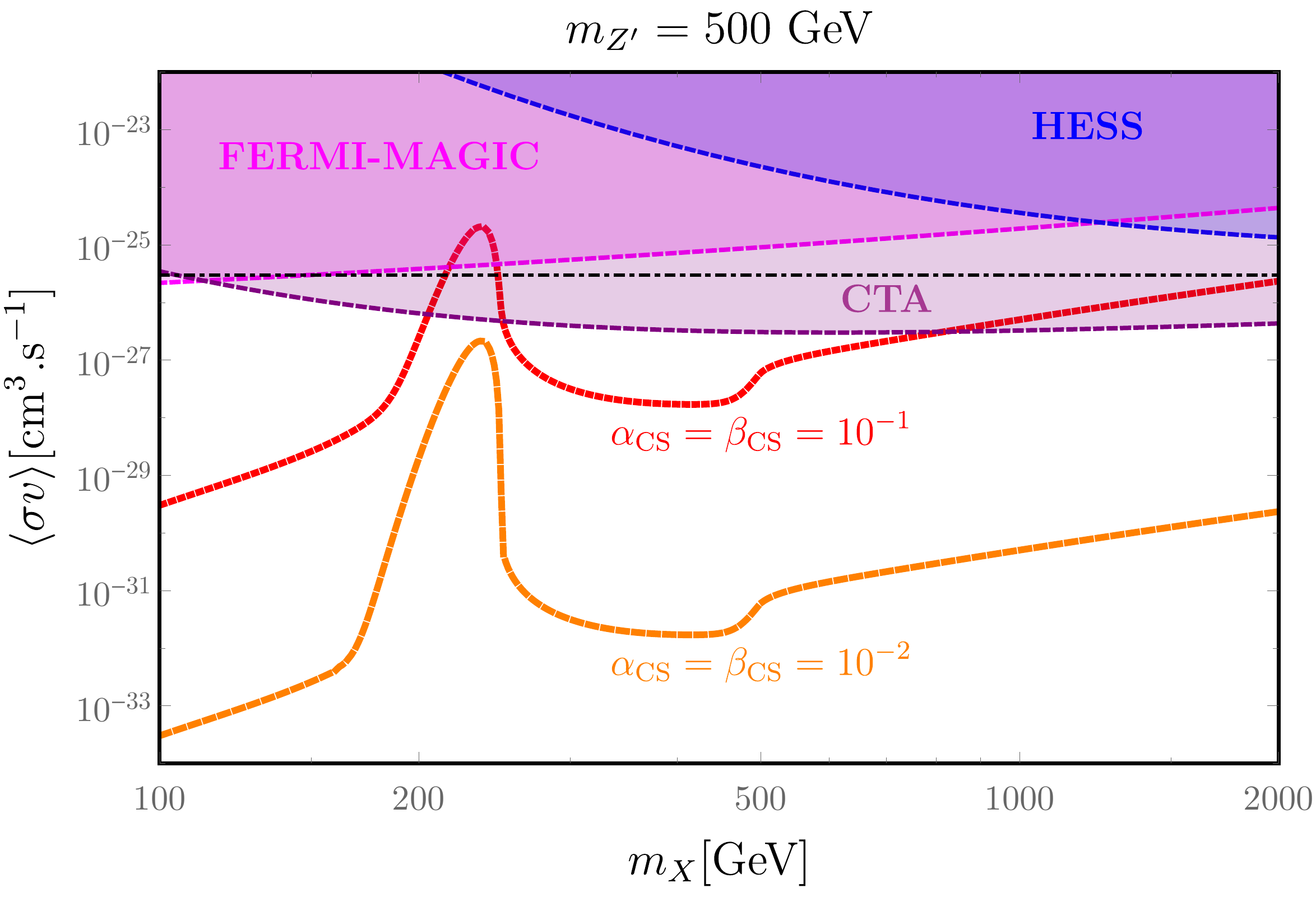}
\caption{Thermally averaged DM pair annihilation cross-section at the time of thermal freeze-out, 
in the presence of double CS terms, as function of the DM mass $m_X$, for $m_{Z'}=500\,\mbox{GeV}$ 
and two assignations for $\alpha_{\rm CS}$ and $\beta_{\rm CS}$, namely $\alpha_{\rm CS}=\beta_{\rm CS}=0.1$ 
(red coloured curve) and $\alpha_{\rm CS}=\beta_{\rm CS}=0.01$ (orange coloured curve). 
The black coloured dashed line represents the thermally 
favoured value of $\langle \sigma v\rangle=$  
$3 \times 10^{-26}\,{\mbox{cm}}^3\, {\mbox{s}}^{-1}$.
Different exclusion regions are the same as of figure \ref{fig:sigmav}.}
\label{fig:sigmav_dCS}
\end{figure}

The behavior of the DM pair annihilation cross-section, as function of its mass, 
is reported in figure~\ref{fig:sigmav_dCS}. Similar to the 
scenario discussed in section \ref{sec:scenario1}, 
the DM pair annihilation cross-sections are typically velocity suppressed except the 
s-wave annihilation channel into $Z'Z'$ final states.
%
%
As evidenced from figure \ref{fig:sigmav_dCS} that value of the thermally averaged DM pair 
annihilation cross-section can match the experimentally favoured value only at  
$m_X \sim \frac{m_{Z'}}{2}$, (i.e., pole region), and for $m_X>m_{Z'}$, so that 
the DM pair annihilation into $Z'$ pairs is allowed. Furthermore, to avoid 
overproduction of the DM, $\alpha_{\rm CS}$ values 
at least $\sim \mathcal{O}(0.1)$ are needed.

\subsection{Indirect detection}

Possible prospects of ID rely, as the kinetic mixing scenario, 
mostly on the detection of gamma-rays produced after the DM pair annihilation into $Z'$ pairs which 
subsequently decay into hadrons. Indeed the other two annihilation channels, i.e., $ZZ$ and $Z\gamma$ have p-wave 
(velocity dependent) annihilation cross-sections  lying several 
orders of magnitude below the present and the near future experimental sensitivities 
\cite{Hooper:2012sr,Gomez-Vargas:2013bea,Ando:2013ff,Gonzalez-Morales:2014eaa,
Queiroz:2014yna,Li:2015kag,Mambrini:2015sia,Massari:2015xea,Queiroz:2016zwd,
Profumo:2016idl,Adams:2016alz,Archambault:2017wyh,Khatun:2017adx,Campos:2017odj}.

That said, similar to the previous case, we assumed that the annihilations into $Z'$ pairs lead to a gamma-ray yield
comparable to the $W^+W^-$ final state. Therefore, one can use CTA sensitivity to DM annihilations into the $W^+W^-$ 
channel to constrain the model as can be seen in figure~\ref{fig:sigmav_dCS}. It is clear that only CTA is expected to 
mildly probe this setup for DM masses above $1$~TeV.

\subsection{Direct detection}

In the case of double CS interactions, the $Z'$ has no direct/tree-level 
couplings with the SM-quarks (or the gluons) and hence, no 
operators relevant for DD is induced at the tree-level.

\subsection{Collider phenomenology}

In the scenario with double CS interactions,
the $Z'$ is directly coupled only with the $Z$-boson and the photon. 
Tree level production of the $Z'$ at the LHC, nevertheless, is possible
through vector boson fusion (VBF) in association with 
two hadronic jets. The production cross-section, however, is more suppressed
compared to the case of a single CS interaction with kinetic mixing
(see section \ref{sec:scenario1})
which has direct couplings with quarks at the tree level.\footnote{A richer collider phenomenology could
appear in extensions of the proposed scenario in which the $Z'$ 
has direct coupling with 
the $W$-boson and the gluons
as studied in refs.
~\cite{Antoniadis:2009ze,Bramante:2011qc,Kumar:2012ba,Dudas:2013sia,Ducu:2015fda}.} 

\begin{figure}[t]
\begin{center}
\includegraphics[width=8.5 cm]{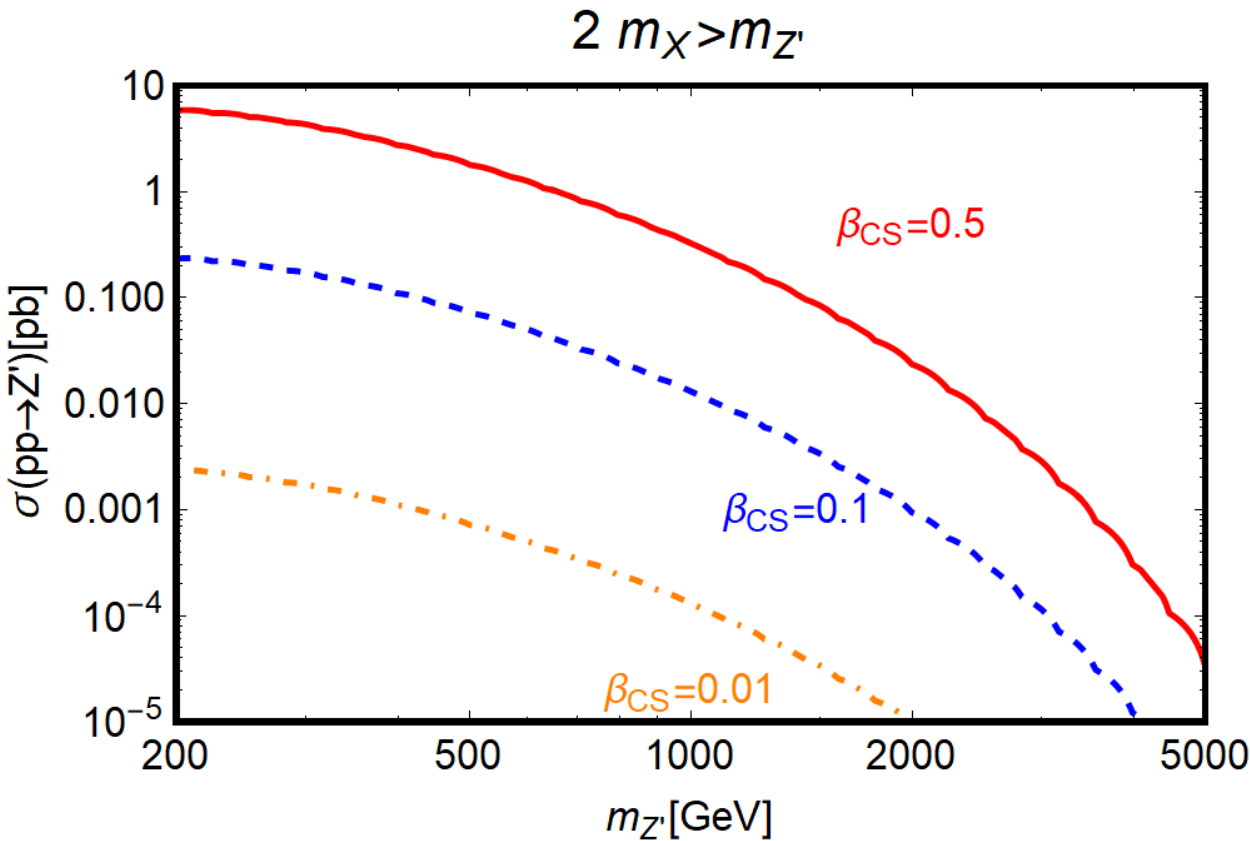}
\end{center}
\caption{Production cross-section of the $Z'$ at the LHC through VBF as 
a function of $m_{Z'}$ for 13 TeV centre-of-mass
energy for three different values of $\beta_{CS}$ parameter, namely, $0.5$, $0.1$ and $0.01$
that are represented with solid red,
dashed blue and dot-dashed orange coloured lines, respectively. 
The choice of $2 m_X > m_{Z'}$ forbids invisible decay of the $Z'$.}
\label{fig:VBF}
\end{figure}

We have reported in figure~\ref{fig:VBF} the expected $Z'$ production cross-section at 
the LHC for 13 TeV centre-of-mass 
energy as a function of the mass of $Z'$ $(m_{Z'})$ for three values of
parameter $\beta_{\rm CS}=0.5,\,0.1$ and $0.01$, depicted with solid red coloured line,
dashed blue coloured line and dot-dashed orange coloured line, respectively. The 
cross-section has been computed using the procedure illustrated in ref.~\cite{Kumar:2007zza}
where discovery prospects of a $Z'$, produced via $Z$-fusion and decaying to four leptons final states,
i.e., $Z' \rightarrow ZZ \rightarrow 4l$, have been investigated.  
These kinds of signals could be probed at the LHC with 14 TeV 
centre-of-mass energy provided that 
$\sigma (pp \rightarrow Z') \simeq 1\,\mbox{pb}$. 
It is now apparent from figure~\ref{fig:VBF} that a future $Z'$ discovery
would appear feasible only for $\beta_{\rm CS}\simeq 0.5$
with $m_{Z'} \lesssim 1$ TeV. For 
lower values of $\beta_{\rm CS}$ the production 
cross-section, goes as $\beta^2_{\rm CS}$, would 
appear very suppressed to escape detection at the LHC unless
(possibly) one considers higher luminosity. One should,
however, remain careful about $\sim \mathcal{O}(1)$ value 
of the parameter $\beta_{\rm CS}$ as this would
indicate a scale for the associated heavy fermion mass $M$ 
(see discussions after eq.~(\ref{eq:twoCSLorigin})) well within the 
reach of ongoing LHC operation where, unfortunately, no evidence
of BSM physics has been confirmed till date.

Notice that in the above discussion 
we have implicitly assumed that invisible decay of the $Z'$ is kinematically forbidden,
i.e., $2\,m_X> m_{Z'}$. 
If this was not the case, the production cross-section of $4l+2j$ final states would be suppressed further by the 
non-zero invisible branching fraction of the $Z'$. On the other hand, a sizable 
invisible branching fraction for $Z'$ might offer meaningful detection
prospects for $p p \rightarrow Z'\rightarrow XX+2j$ process. Investigation 
of such signature would require a dedicated study which 
is beyond the scope of this work.

\subsection{Results}
\begin{figure}[t!]
\centering
\includegraphics[width=7.6cm]{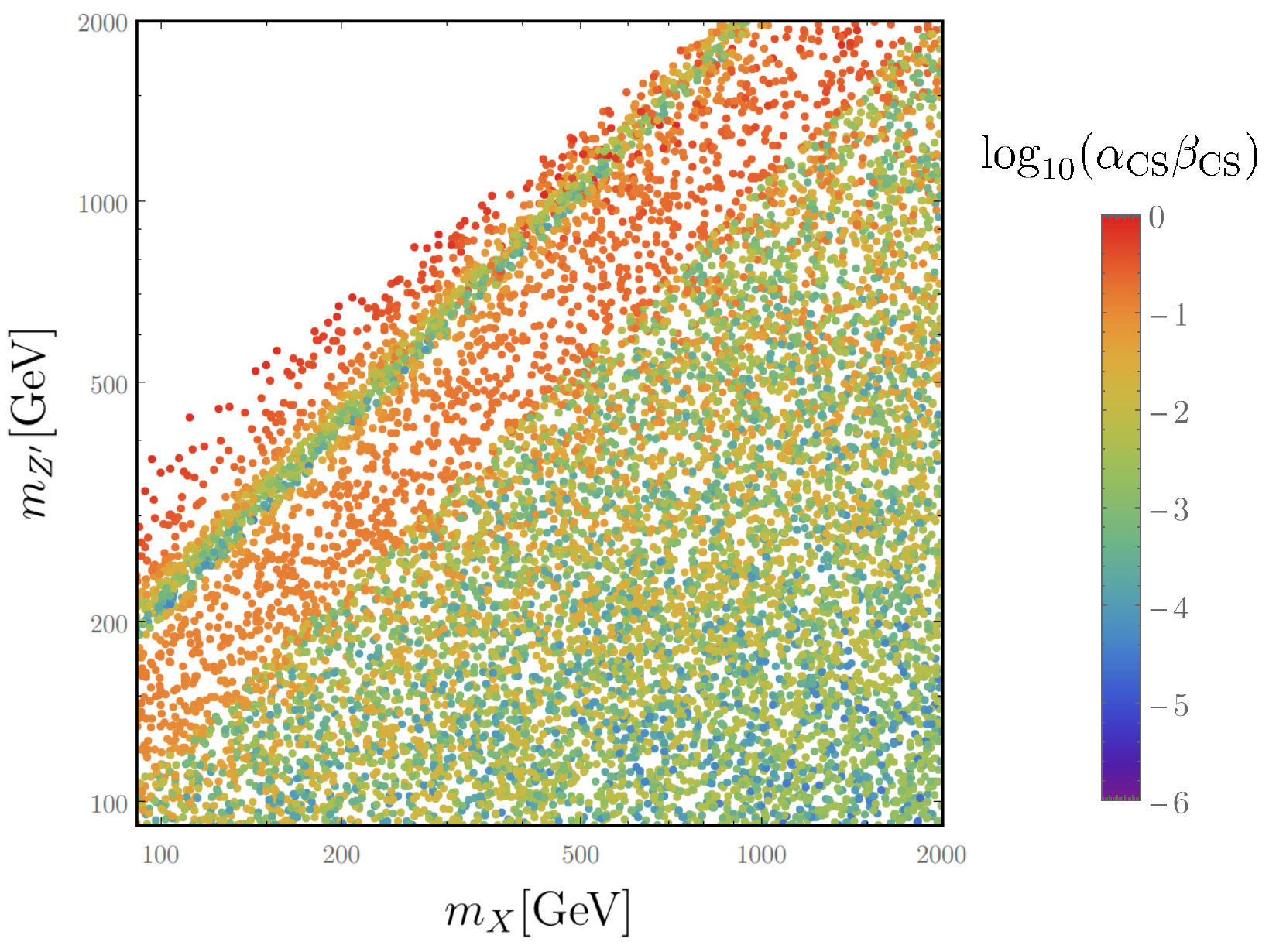}
\includegraphics[width=7.0cm]{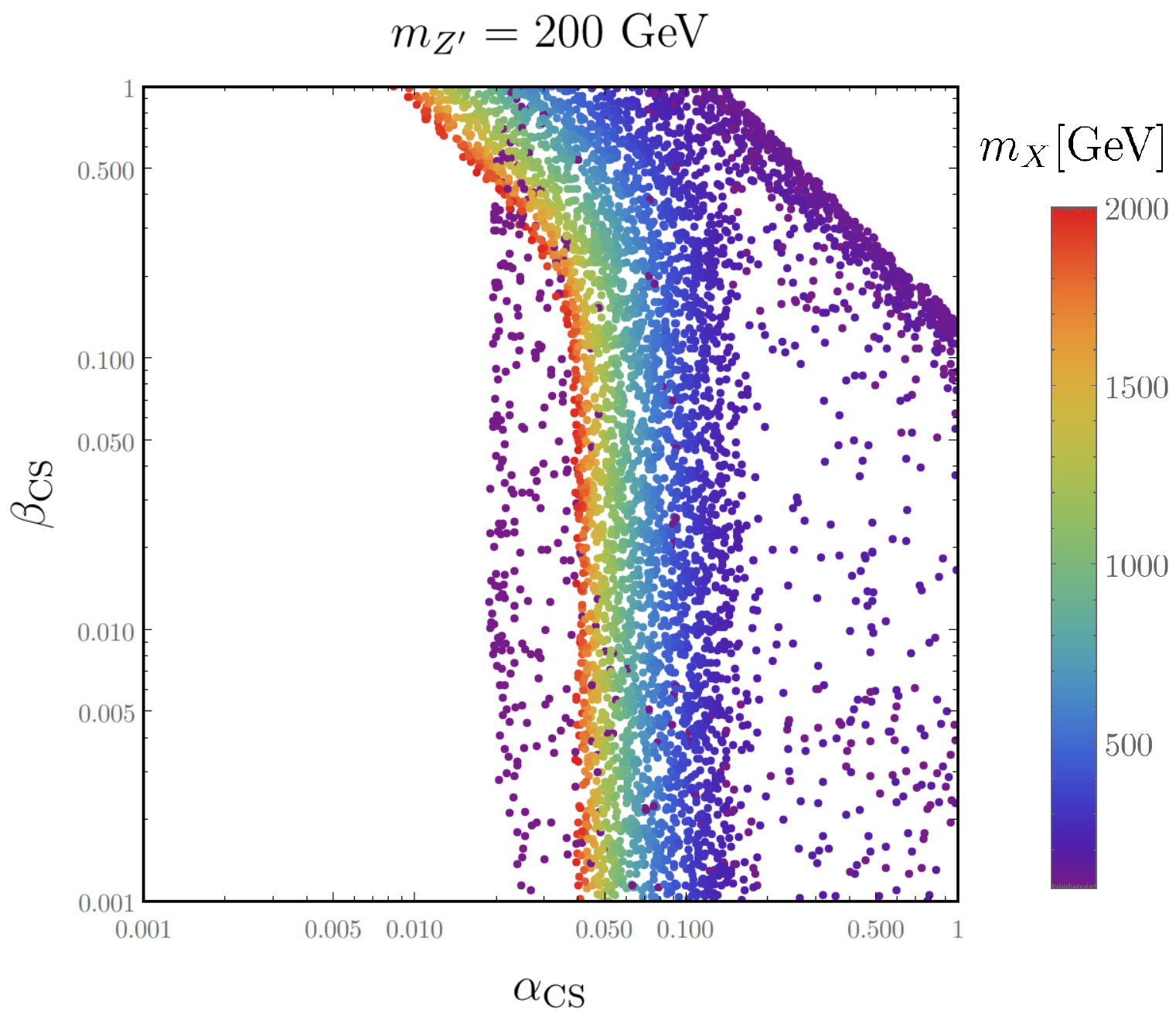}
\caption{Results of a parameter scan for the setup with two CS terms
showing points compatible with the DM phenomenology.
The left plot is in $m_X,\,m_{Z'}$ bi-dimensional plane with
a colour coding for the product of two CS couplings
$\alpha_{\rm CS}\beta_{\rm CS}$ while the right plot is in 
$\alpha_{\rm CS},\,\beta_{\rm CS}$ plane for $m_{Z'}=200$ GeV with a 
colour coding for various $m_X$ values.}
\label{fig:scandoubleCS}
\end{figure}

Similar to the first model considered in this work, we perform a scan in 
the parameter space over $\alpha_\text{CS},\,\beta_\text{CS},\,m_{Z^\prime}$ and $m_X$ 
and represent our findings in figure~\ref{fig:scandoubleCS} showing the phenomenologically viable model points.
One should note that, contrary to the kinetic mixing
scenario, the absence of tree-level $Z'$ couplings
with the SM fermions appears useful to efface a set of constraints
coming from the precision $Z$-physics and collider observations.
The scan is performed in the following ranges:

\beq
\label{eq:scan2}
\alpha_{\rm CS} \in \left[10^{-3}, 1\right],\,\,\,\,
\beta_{\rm CS} \in \left[10^{-3}, 1\right],\,\,\,\,
m_X \in \left[90 \,\mbox{GeV},2\,\mbox{TeV}\right],\,\,\,\,
m_{Z^\prime} \in \left[90\,\mbox{GeV},2\,\mbox{TeV}\right].
\eeq
%

As expected, the left panel of figure~\ref{fig:scandoubleCS}, representing the viable model 
points in the bi-dimensional plane $(m_X,m_{Z'})$, shows, similar to the kinetic mixing scenario, 
a sensitive preference for configurations with $m_X > m_{Z'}$, for which the correct relic density 
can be more easily achieved through the velocity unsuppressed annihilation rate into $Z'$ pairs. This is 
further evidenced by the right panel of figure~\ref{fig:scandoubleCS}, investigating the bi-dimensional 
plane $(\alpha_{\rm CS},\,\beta_{\rm CS}$) for a fixed $m_{Z'}$. Indeed, most of the points approximately trace isocontours 
with a shape independent of $\beta_{\rm CS}$, unless this coupling 
is much bigger than $\alpha_{\rm CS}$. 
This behavior would be exactly expected in the case when the DM relic density is mostly accounted by the 
annihilation into the $Z'Z'$ final state since the corresponding rate depends only on $\alpha_{\rm CS}$ (see eq.~(\ref{eq:XXzpzp2})). 
One must note that the demand of $\beta_{\rm CS} \gg \alpha_{\rm CS}$ is theoretically
challenging since normally one would expect $\beta_{\rm CS} < \alpha_{\rm CS}$ as
already discussed in the context of eq.~(\ref{eq:twoCSLorigin}).

It is curious to note the apparent conflicts among the DM detection prospects
in DD, ID and collider experiments with $Z'$ searches at the collider and
theoretical consistency of the studied framework.
For example, from figure \ref{fig:VBF} it is evident
that a $Z'$ discovery at the LHC with 13 TeV centre-of-mass
energy would require $\beta_{\rm CS}\sim \mathcal{O}(1)$ and $m_{Z'}\lesssim 1$ TeV.
Such high value of $\beta_{\rm CS}$, as already mentioned,
is hard to accommodate phenomenologically. Pushing $\beta_{\rm CS}$
towards its natural regime, i.e., $\sim \mathcal{O}(10^{-4})$, on the other hand, would predict a 
$\alpha_{\rm CS}\sim \mathcal{O}[10-10^4]$ to retain some of the viable 
points, notably for log$_{10}(\alpha_{\rm CS}\beta_{\rm CS})\gtrsim {-3}$, as
shown in the left plot of figure \ref{fig:scandoubleCS}
to respect the DD, ID and relic density constraints. Now keeping
in mind the radiative origin of $\alpha_{\rm CS}$, as explained
in the appendix \ref{sec:appendixC}, such high values of $\alpha_{\rm CS}$
would require either a pathologically behaved strongly
coupled theory or unnatural high values for the associated charges.
Lower and rather natural $\alpha_{\rm CS}$ values, e.g., $\sim 10^{-3}$,
can ameliorate such theoretical shortcomings and one can still
get viable points consistent with the DM observables
in the region of log$_{10}(\alpha_{\rm CS}\beta_{\rm CS})\lesssim {-5}$.
The associated low $\beta_{\rm CS}$ values, however, failed
to produce detectable $Z'$ signals at the LHC as already shown in
figure \ref{fig:VBF}. Similar contradictions also appear
for the right plot of figure \ref{fig:scandoubleCS} which,
keeping in mind the collider detection prospect of a $Z'$, 
predicts $0.01 \lesssim \alpha_{\rm CS}\lesssim 0.3$ for $m_{Z'}=200$ GeV.
This range of the parameter $\alpha_{\rm CS}$ once again
either requires moderate values of the involved charges
or asks for a strongly coupled model frameworks.
In a nutshell, keeping in mind the detection possibilities
of the BSM physics, either in DM or in collider searches,
together with an elegant theoretical construction, the scenario with two
CS interactions is less appealing compared
to a scenario with one CS interaction and a kinetic mixing.
Such simple conclusion, however, is not obvious when
a scenario with two CS interactions involves
non-Abelian gauge groups.

\section{Origin of a Chern-Simons coupling and UV completion }
\label{sec:UVcompletion}
In this section we propose a UV complete model that 
can address the origin of CS and  kinetic mixing terms.
The generation of a generalized CS interaction term between the 
gauge boson of a BSM $U(1)$ group and the SM gauge bosons has already been
proposed in refs.~\cite{Anastasopoulos:2006cz,Dudas:2009uq,Antoniadis:2009ze}. 
The effective CS coupling is originated from the triangle loops 
involving BSM heavy fermions, charged under 
both the SM $U(1)_Y$ and a BSM $U(1)$ symmetry groups,
after integrating out the heavy degrees of freedom.
In this work we consider a similar construction where a CS 
interaction term is generated between the gauge bosons of two new BSM symmetry groups 
labeled as $U(1)_X$ (associated to the DM) and $U(1)_V$ (associated with $\wt{V}_\mu$), respectively.\footnote{For the convenience of
reading note that in scenario-I, $\wt{V}_\mu$ denotes the gauge field of $U(1)_V$ which, after 
diagonalization of the flavour basis Lagrangian, is written as $Z'_\mu$ while
in scenario-II the kinetic and mass terms are already diagonalized
and hence, $\wt{V}_\mu=Z'_\mu$.}

Concerning the origin of kinetic mixing, in the most general scenario, i.e., when the new BSM 
fermions are charged\footnote{One could expect the BSM fermions to have an impact on the 
DM decoupling process from the SM thermal bath, however, since the freeze out occurs only 
when the DM particles are non-relativistic, the
BSM fermions would have already decoupled and their density would be exponentially 
suppressed due to Boltzmann factor.} under the BSM gauge groups $U(1)_X,\, U(1)_V$ as well as
with respect to $U(1)_Y$ of the SM, one can radiatively generate three possible kinetic mixing
terms like $X^{\mu\nu}\wt V_{\mu\nu}, \, X^{\mu\nu} B_{\mu\nu}$ and
$\wt V^{\mu\nu} B_{\mu\nu}$ with\footnote{Such terms, being Lorentz invariant, renormalizable 
and invariant under the SM gauge groups, can directly appear in  
Lagrangian. We, however, do not consider such possibility
and confine our discussion on scenarios where such interactions are radiatively generated.} 
appropriate field strengths of the involved
Abelian groups. The co-efficients for these terms emerge as a result
of integrating out the aforesaid heavy BSM fermionic degrees of freedom 
and also include associated gauge charges. One should remain
careful at this stage as a kinetic mixing like $ X^{\mu\nu} B_{\mu\nu}$
triggers a mixing between the DM $X_\mu$ and $Z_\mu$ which
would subsequently allow the DM to decay into the SM fermions
and thereby, spoiling its stability. A similar situation can also
appear for $X^{\mu\nu}\wt V_{\mu\nu}$, given that $\wt V_\mu$
is somehow, e.g., via a second kinetic mixing $\wt V^{\mu\nu} B_{\mu\nu}$, 
decaying into the SM particles. It is thus important to assign
gauge charges for the BSM fermions in an elegant way such that
the stability of the DM remains preserved.

We denote new BSM fermions as $\chi$ ($\psi$), chiral with respect to $U(1)_X$ ($U(1)_V$)
while vector like compared to $U(1)_V$ ($U(1)_X$) as well as to the SM \cite{Antoniadis:2009ze}.
This way no new chiral anomalies are introduced in the SM while one can find 
suitable charge assignments for these BSM fermions to efface new anomalies.
Regarding the mass generation of these new fermions the most plausible option is represented via 
Yukawa interactions with two BSM complex scalar fields $\phi_{X,V}$. The 
relevant Lagrangian is written as:
%
\bea
\label{eq:bsmfm}
\mathcal{L}^\text{fermions}=- y_F \phi_V \ovl{\psi}_{1L} \psi_{1R}- y_F \phi_V^* \ovl{\psi}_{2L} \psi_{2R} -  y_F \phi_X \ovl{\chi}_{1L} \chi_{1R}- y_F \phi_X^* \ovl{\chi}_{2L} \chi_{2R} + \text{h.c.},
\eea
where $y_F$ represents generic Yukawa couplings.\footnote{We, without
the loss of generality, consider identical Yukawa couplings for all the BSM 
fermions for simplicity since the final results are independent of them.}
The complex scalar fields $\phi_X$, $\phi_V$ can be written using the 
following parametrization as:
\begin{equation}
\phi_X=(v_X+h_X)e^{i\theta_X/v_X} \qquad \text{and} \qquad \phi_V=(v_V+h_V)e^{i\theta_V/v_V},
\end{equation}
with $v_X,\,v_V$ and $h_X,\,h_V$ denoting the corresponding VEVs
and the Higgs fields. Here $\theta_X,\,\theta_V$
are Stueckelberg axions \cite{Stueckelberg:1900zz,Stueckelberg:1938zz,Kors:2004dx}. Investigating the kinetic
terms for these scalars with suitable BSM covariant derivatives,
one gets, for example, for $\phi_X$:

\begin{equation}
|D_\mu \phi_X|^2 = |\partial_\mu \phi_X -i q_X g_X \phi_X X_\mu|^2\supset 
(\partial_\mu \theta_X - q_X g_X v_X X_\mu)^2, 
\end{equation}
with $g_{X}$ as the gauge coupling of the $U(1)_X$ gauge group
and $q_X$ as the charge of $\phi_X$ with respect to this group. 
The DM mass thus, is generated via the 
Stueckelberg mechanism \cite{Stueckelberg:1900zz,Stueckelberg:1938zz,Kors:2004dx}. The 
masses for the new BSM fermions, $\chi_1,\,\chi_2$, are also generated after the spontaneous symmetry
breaking (SSB) in the $U(1)_X$ sector.
In fact, after SSB one ends up with the following tree-level
realizations for various masses:
\begin{equation}
\label{eq:XVgenmass}
m_{h_X}\sim \sqrt{\lambda_X} v_X, \qquad m_{F} \sim y_F v_X, \qquad m_X \sim g_X q_X v_X,
\end{equation}
where $\lambda_{X}$ is the quartic coupling of 
the $\phi_{X}$ potential and the notation $m_F$ is used to represent
generic BSM fermion masses. A similar construction holds for 
$U(1)_V$ with $\lambda_V,\,g_V,\,q_V,\,m_{h_V}$ and $m_V$ as the appropriate replacements.
We further consider the following hierarchy among the various couplings: 
$\lambda_X,\,\lambda_V \gg y_F \gg g_X,\,g_V$. Such choice implies
$m_{h_X},m_{h_V} \gg m_{F} \gg m_X,\,m_V$ for $\lambda_X,g_X\sim \lambda_V,\,g_V$
with $v_X\sim v_V$.
Hence, at an energy scale $E\sim m_F \ll m_{h_X},\,m_{h_V}$, the new scalars
$h_X,\,h_V$ are nearly decoupled from the theory given that 
$v_X,\,v_V$ are reasonably high. In this limit one can use the following
approximations $\phi_X \simeq v_X e^{i\theta_X/v_X}\simeq v_X+i\theta_X$ and 
$\phi_V \simeq v_V e^{i\theta_V/v_V}\simeq v_V+i\theta_V$ which help
to recast eq.~({\ref{eq:bsmfm}}) as:
%
\begin{equation}
\mathcal{L}^\text{fermions} \supset - i y_F \theta_V \ovl{\psi}_1 \gamma_5 \psi_1 + i y_F \theta_V \ovl{\psi}_2 \gamma_5 \psi_2 - i y_F \theta_X \ovl{\chi}_1 \gamma_5 \chi_1 + i y_F \theta_X \ovl{\chi}_2 \gamma_5 \chi_2.
\end{equation}

\begin{table}[t]
\begin{center}
\begin{tabular}{|c||c|c|c|c||c|c|c|c|}
\hline 
 & $\psi_{1L}$ & $\psi_{1R}$ & $\psi_{2L}$ & $\psi_{2R}$ & $\chi_{1L}$ & $\chi_{1R}$ & $\chi_{2L}$ & $\chi_{2R}$ \\ 
\hline 
\hline
$U(1)_X$ & $e_1$ & $e_1$ & $e_2$ & $e_2$ & $e_4$ & $e_3$ & $e_3$ & $e_4$ \\ 
\hline 
$U(1)_V$ & $q_1$ & $-q_1$ & $-q_1$ & $q_1$ & $q_2$ & $q_2$ & $-q_2$ & $-q_2$ \\ 
\hline 
\end{tabular} 
\end{center}
\caption{Charge assignments of the left- and right- chiral components of 
the BSM fermions which belong to gauge group $U(1)_{X}\times U(1)_V$.}
\label{tab:charges}
\end{table}

The structure of Yukawa couplings of the new fermions,
as well as 
the requirement of anomaly cancellations, restrict the possible charge assignments 
under the BSM $U(1)$ groups. 
A set of assignations complying with these two requirements is reported in 
table~\ref{tab:charges}. 
As evident from table~\ref{tab:charges} that this kind of charge assignations 
allows natural cancellations of the $U(1)_{V,X}^3$ 
anomalies, independently for the $\psi$ and $\chi$ sectors, irrespective of the values 
of $q_i,e_i$ charges.\footnote{One could see from table \ref{tab:charges} that the sum
of all left- and right-chiral charges for $\psi$ and $\chi$
vanishes for $U(1)_V$ while for $U(1)_X$ it is $2\sum\limits^4_{i=1}e_i$.
If we also set this sum to be zero, as expected from the requirement
of gauge-gravity anomaly cancellation for an Abelian group, we can
recast eq.~(\ref{eq:anorel}) as $q_2=\frac{q_1(e_2-e_1)}{e_3-e_4}$.}
The cancellation of the mixed anomalies, e.g.,
$U(1)_V \, U(1)^2_X$, is instead achieved in a non-trivial way. 
This, indeed, requires the following relation between the charges:
%
\begin{equation}
\label{eq:anorel}
q_2=\frac{q_1 (e_1^2-e_2^2)}{ (e_3^2-e_4^2)}.
\end{equation}

Further, the charge assignments of table~\ref{tab:charges} 
predicts a vanishing kinetic mixing between $X$ and $V$ as 
\begin{equation}
\label{eq:XVkinmix}
\sum\limits^{i=1,\,2}_{\xi=\psi_i,\,\chi_i} 
c^X_{\xi_L} c^V_{\xi_L}+ c^X_{\xi_R} c^V_{\xi_R}=0,
\end{equation}
with $c^{X(V)}_{\xi_{L(R)}}$ representing appropriate charges
shown in the table~\ref{tab:charges}. This, as discussed already, is crucial since
a mixing between $X$ and $V$, in the presence
of a kinetic mixing between $V,\,Z$, can trigger
a subsequent mixing between $X$ and $Z$, such that
$X$ can decay into the SM fermions and thereby, 
the stability of the DM gets spoiled.

We show in detail later in appendix \ref{sec:appendixC} that when eq.~(\ref{eq:anorel})
is satisfied, it appears feasible to construct an anomaly
free theory where the following effective operator emerges
after integrating out heavy fermionic degrees of freedom
from the triangular loops:
%
\begin{equation}
\label{eq:sasaX}
\epsilon^{\mu \nu \rho \sigma} D_\mu \theta_X D_\nu \theta_V X_{\rho \sigma},
\end{equation}
where $\theta_X,\,\theta_V$ are Stueckelberg axions
of the $U(1)_X,\,U(1)_V$ groups and $D_\mu\theta_X = \partial_\mu \theta_X-g_X q_X v_X X_\mu$, and 
$D_\nu\theta_V = \partial_\nu \theta_V-g_V q_V v_V \wt V_\nu$ with 
$X_\mu,\,\wt V_\nu$ as gauge bosons of the concerned
$U(1)_X,\, U(1)_V$ groups, respectively. Eq.~(\ref{eq:sasaX}) is invariant under the following gauge transformations:
\begin{equation}
 X_\mu \rightarrow X_\mu + \partial_\mu \alpha_X, \,\,\, \wt V_\mu \rightarrow \wt V_\mu 
 + \partial_\mu \alpha_V, \,\,\, \theta_X \rightarrow \theta_X + g_X q_X v_X \alpha_X, \,\,\,
 \theta_V \rightarrow \theta_V + g_V q_V v_V \alpha_V,
\end{equation}
with $\alpha_X$ and $\alpha_V$ as the transformation parameters. The same equation, after considering unitary gauge, 
leads to the following operator
as introduced earlier in eq.~(\ref{eq:starting_lagrangian}):
\begin{equation}
\mathcal{L} = \alpha_\text{CS} \epsilon^{\mu \nu \rho \sigma} X_\mu \wt V_\nu X_{\rho \sigma}, 
\end{equation}
with
\begin{equation}
\label{eq:alphaCSwork}
\alpha_\text{CS} \equiv \frac{q_1 \left(e_2^2-e_1^2\right)}{8 \pi ^2}.
\end{equation}

One can define an effective charge $\wt{Q}^3\equiv q_1(e_2^2-e_1^2)$ 
to get a simple relation:
%
\begin{equation}
\alpha_{\rm CS} =\dfrac{\wt{Q}^3}{8\pi^2}.
\end{equation}

Clearly $\alpha_{\rm CS}$ $\sim \mathcal{O}$ $[ 10^{-2},1]$
(see eq.~(\ref{eq:scan1})) or $\sim\mathcal{O}$ $[ 10^{-3},1]$ 
(see eq.~(\ref{eq:scan2})) corresponds to a $\sim \mathcal{O}(1)$ value of the effective charge $\wt Q$ for 
one generation of the BSM fermions.

Note that the CS coupling $\alpha_{\rm CS}$ has no explicit dependence
on the BSM fermion mass, i.e., it seems to remain finite as
$m_F$, the relevant heavy fermion mass, $\longrightarrow \infty$ and thereby, 
resembles a non-decoupling effect. This is a consequence of the assumption
$\lambda_{X},\,\lambda_{V} \gg y_F \gg g_X,\,g_V$, as considered
earlier, which makes $\alpha_{\rm CS}$ independent of $m_F$
as long as $m_F \gg m_X,\,m_{Z'}$ such that the adopted effective
approach remains justified for an energy scale $E$ below 
the mass of the ``lightest'' BSM fermion of the theory.
The parameter $\beta_{\rm CS}$ (see eq.~(\ref{eq:twoCSLorigin})),
on the contrary, vanishes as the associated BSM fermion masses
$\longrightarrow \infty$, as expected according to the decoupling effect.

One should further note that as the effective charge $\wt Q$ includes the gauge couplings in its definition,
$\wt{Q}\sim \mathcal{O}(1)$ implies either $g_X\sim g_V\sim \mathcal{O}(1)$ or a large multiplicity of the BSM 
fermions having gauge charges $\sim \mathcal{O}(1)$.
However, as mentioned  previously, the aforementioned
theoretical construction relies on the assumption of $\lambda_{X},\,\lambda_{V} \gg g_X,\,g_V$ which,
for $g_X,\,g_V\sim \mathcal{O}(1)$,
hints towards a strongly coupled theory. In this regime one would encounter several
theoretical issues like the vacuum instability, etc. which might spoil viability of the
effective approach. Further discussions of such issues are beyond the theme of this paper and 
we note in passing that from the view point of a radiative origin, 
$\alpha_{\rm CS} \sim \mathcal{O}(1)$ is unnatural.

The kinetic mixing parameter $\delta$ (see eq.~(\ref{eq:starting_lagrangian})), as already stated 
in the beginning of this section, can get generated at the loop level
from two sets of the BSM fermions (preferably vector-like to avoid new anomalies in a trivial way)
charged under the SM $U(1)_Y$ and BSM $U(1)_V$ groups and having masses
$M_y$ and $M_v$, respectively. The parameter $\delta$ is then estimated
as $\simeq$ $(q_Y g_Y q_V g_V/16\pi^2)\times {\rm log}(M_v/M_y)$ \cite{Holdom:1985ag} with 
$q_Y,\,q_V,\,g_Y,\,g_V$ as the relevant combination of gauge charges and gauge
couplings of the associated gauge groups. Assuming these gauge charges
and couplings, as well as log$(M_v/M_y)$, to be $\sim \mathcal{O}(1)$, 
one would expect natural range of $\delta$ as $\sim \mathcal{O}(10^{-3}-10^{-2})$.
This range, as evident from eq.~(\ref{eq:alphaCSwork}) and eq.~(\ref{eq:effctivealphaCS}), 
is almost the same as of $\alpha_{\rm CS}$, considering
$\sim \mathcal{O}(1)$ values of the involved gauge charges. Both
these natural ranges of CS coupling $\alpha_{\rm CS}$ and 
kinetic mixing parameter $\delta$ are connected with their
radiative origins.

\section{Summary and conclusions}
\label{sec:Conclusion}

In this article we scrutinized experimental viability and theoretical consistency of the WIMP
DM models comprised of an Abelian vectorial DM $X_\mu$ and an Abelian $Z'$ portal, coupled
through a CS interaction. We studied two possibilities of connecting
the $Z'$ with the SM: (1) via a kinetic mixing with the SM hypercharge
field strength and, (2) another CS coupling with $Z$ boson
and field strength of the SM $U(1)_Y$ group. We successively
addressed the DM and collider phenomenologies of these two scenarios.
Regarding the DM phenomenologies we investigated the detection
prospects of these two frameworks in the light of accommodating the correct relic density
and sensitivity reaches of the various existing as well as
anticipated upcoming DD and ID experiments. Concerning collider probes we examined
the observational aspects of these models from the view point
of dijet, dilepton resonances and mono-{\bf X} searches
using the 13 TeV LHC data. Further, we also studied the 
viable ranges of the associated parameters focusing on
the possible theoretical and/or ``well-measured'' experimental constraints,
mainly for the kinetic mixing scenario, arising from the 
EWPT, $\rho$-parameter, $Z$-mass, total and invisible $Z$-decay widths, etc.
Finally, we also explored possible origins of a kinetic mixing term
and a CS interaction term, arising via a set of heavy BSM fermions running
in the triangle loops, from the standpoint of an UV complete
theory. These fermions
are charged under the new BSM gauge groups, connected with the 
Abelian DM and the $Z'$, and possess non-trivial charges
(or vector-like) with respect to the SM.  We explained how a specific
choice of the associated charges for these
fermions can appear instrumental to acquire a CS coupling, a
kinetic mixing term for $Z$-$Z'$, an anomaly free model setup
and the DM stability.

Given that $X_\mu$ is the ``only'' DM 
candidate, one observes that the DM experimental detection/exclusion potentials
(correct relic density and sensitivities towards DD, ID experiments),
for the model with one CS coupling and a kinetic mixing,
appear promising for $m_{Z'}$ preferably below $1$ TeV with
$m_X \geq m_{Z'}$ configuration, expect the pole region (i.e., $2m_X \backsimeq m_{Z'}$),
for $\delta \sim \mathcal{O}(0.1),\, \mathcal{O}(0.1) \lesssim
\alpha_{\rm CS} \sim \mathcal{O}(1)$. Such ranges for $\delta,\,\alpha_{\rm CS}$ parameters
as well as low $m_{Z'}$ are noticed to be
incompatible with the collider searches of heavy dilepton
resonances and mono-{\bf X} events. One can, of course, lower
the value of parameter $\delta \sim \mathcal{O}(10^{-3})$ to efface
the existing collider constraints. Such values of $\delta$ 
would predict $\lesssim \mathcal{O}(1)$ events even if
one considers high-luminosity run of the LHC or future
proton-proton colliders with higher centre-of-mass energies
and thus, conceal the model from collider searches till a faraway future. As a quantitative example, to get $\sim \mathcal{O}(10)$ events 
for TeV scale new physics, i.e., $m_{Z'}=1$ TeV, from $pp\to Z'\to jj,\,l^+l^-$ processes at the 100 TeV proton-proton collider,
assuming a ``$100\%$'' detection efficiency, one would require
$100~{\rm fb}^{-1}$ of integrated luminosity to probe the $\delta=10^{-3}$ configuration. The situation
is similarly disappointing concerning the DM phenomenology with
such $\delta$ values, unless sensitivity reaches of the relevant
DD, ID experiments are improved by several orders of magnitude
and the idea of multi-component DM is invoked to account for the 
correct relic density. One should note that $\delta \lesssim 10^{-3}$
and hence, the resultant suppressed $Z$-$Z'$ mixing, can
peacefully co-exist with the constraints of EWPT, $\rho$-parameter
and the measured $Z$-boson mass and decay (total, invisible\footnote{Depends
also on $\alpha_{\rm CS}$ and irrelevant for $2m_X > m_Z$.}) widths.
In a nutshell, we see that $\delta$ value in the ball park of $10^{-3}$ 
appears experimentally unpleasant. Nevertheless, assigning
a radiative origin to this coupling, involving heavy BSM fermions,
$\delta\lesssim \mathcal{O}(10^{-3})$ is the natural range that one would expect for the parameter
$\delta$, unless a high multiplicity of the associated
BSM fermions is considered to raise this parameter (at least) by
one order of magnitude that makes the chosen setup testable at
the ongoing and upcoming experiments. For example, with $\delta=0.01$, a $1$ TeV $Z'$ can
produce $\sim\mathcal{O}(10)$ ``detectable'' events in dijet/dilepton resonance searches itself at the 13 TeV 
LHC with about $100~{\rm fb}^{-1}$
of integrated luminosity, even assuming a $``10\%"$ detection efficiency.
A radiative origin for CS coupling $\alpha_{\rm CS}$,
from the perspective of an UV complete construction, also predicts
a natural range for $\alpha_{\rm CS}$ as $\lesssim \mathcal{O}(10^{-3})$
whereas discovery/exclusion prospects, with the existing and 
near future experimental setups, favour $\alpha_{\rm CS}\sim \mathcal{O}(1)$.
Such $\alpha_{\rm CS}$ values, along with a $\delta$ of similar order,
can accommodate the correct relic density rather easily and can be probed/excluded
form DD, ID and collider (via mono-{\bf X}) searches.
An $\mathcal{O}(1)$ value of $\alpha_{\rm CS}$, just like 
the kinetic mixing parameter $\delta$, is hard to explain 
with a radiative origin unless more families of the BSM fermions
are included. Any such non-minimal constructions, i.e., large 
number of BSM fermions to increase $\delta$ and/or
$\alpha_{\rm CS}$ value(s) or a multi-component
DM to account for the correct relic density with ``natural'' $\delta$
values would reduce the model predictivity. 
Moreover, the underlying assumptions behind 
the UV complete construction of a CS coupling can easily lead 
to a strongly coupled theory where the validity of the adopted
effective approach might seem questionable.

Experimental attainments of the second case study with two 
CS couplings are more contrived due to the absence of a tree-level
mixing between the $Z'$ and the SM fermions, unlike the first
case study with one CS coupling and a kinetic mixing.
Missing tree-level couplings between $Z'$ and 
the SM fermions conceal this framework from constraints
like the EWPT, $\rho$-parameter, precision $Z$ physics, etc.
which offer notable effects on the model parameter space for scenario-I.
In this framework, DD prospects are missing at the tree-level
and ID sensitivities remain orders of magnitude below
the ongoing and upcoming experimental reaches. The requirement
of the correct relic density, except the pole region,
is possible for $m_X > m_{Z'}$ with $\alpha_{\rm CS},\,\beta_{\rm CS}$
$\gtrsim \mathcal{O}(0.1)$ even at low $m_{Z'}$. This regime
of the two CS coupling values is resourceful for a LHC
detection, although the high-luminosity run appears to be the preferred choice.
The collider discovery/exclusion potentials are primarily sensitive
to $\beta_{\rm CS}$ except when one considers 
invisible decay of the $Z'$ (i.e., mono-{\bf X} searches) where $\alpha_{\rm CS}$ also enters
in the analysis. The experimentally favoured ranges
of the two CS coupling values, just like the kinetic mixing scenario,
are hard to realize from their respective theoretical origins.
The conclusion for $\alpha_{\rm CS}$ remains the same
as of the kinetic mixing scenario, since it  possesses the same
radiative origin involving heavy BSM fermions from the standpoint of an
UV complete construction. The origin of the second CS coupling $\beta_{\rm CS}$
is also connected to the BSM fermions, however, in 
a more convoluted way that allows a vanishing value for $\beta_{\rm CS}$
as the associated fermions masses get heavier and decouple from the 
low-energy theory. Further, by construction, in general one expects $\alpha_{\rm CS} > \beta_{\rm CS}$
which indicates natural range of $\beta_{\rm CS}$ in the ballpark of 
$10^{-4}$. Thus, this scenario remains practically hidden 
from the collider searches, even considering the high-luminosity LHC
or a $100$ TeV proton-proton collider. The latter, being quantitative, 
would require a $1000~{\rm fb}^{-1}$ of integrated luminosity
to yield $\sim \mathcal{O}(1)$ events for a $1$ TeV $Z'$ with $\beta_{\rm CS}=10^{-4}$,
assuming a ``$100\%$'' detection efficiency. The requirement of the 
correct relic density also faces similar hindrance. The trick of pushing $\beta_{\rm CS}$
values upwards by adding more 
BSM fermions is rather intricate compared to the kinetic mixing
scenario. 

The key outcomes of this study, based on the obtained results, 
give the following conclusions: (1) scenario-II with two CS couplings,
both connected to some Abelian groups, would appear rather hard to
probe experimentally even in the near future experiments
if one sticks to a theoretically ``well-behaved'' construction. 
Moving to a setup where the second CS coupling is connected
to some non-Abelian groups, for example, bridging an Abelian $Z'$ with
gluons, one can efface such limitations and 
the concerned scenario could produce detectable signals
at the DM and the collider experiments. (2) Phenomenological
viability of scenario-I with one CS coupling and a kinetic
mixing, together with an UV complete model construction, also faces challenges to comply with the relevant
experimental search sensitivities, however, in a lessened
way. In particular, $\mathcal{O}(10^{-3}) \lesssim \delta \lesssim \mathcal{O}(10^{-2})$ configuration
with $m_{Z'}$ around a TeV could be probed at the 14 TeV LHC
with full luminosity as well as in high-luminosity
LHC or in the next generation proton-proton colliders with increased
centre-of-mass energies. The DD, ID prospects, with naturally preferred $\alpha_{\rm CS}$ values, however, still
remain well beyond the reaches of the next generation experiments
although the requirement of the correct relic density can be accounted
for by adding other DM candidates.

\section*{Acknowledgements}
The authors would like to thank A. Pukhov and B. Fuks for substantial help with  MicrOmegas and FeynRules, E. Dudas, G. Bhattacharyya, H.M. Lee and H. Murayama,  for enlightening discussions. P.~G. acknowledges the support from P2IO 
Excellence Laboratory (LABEX). This work is also supported by the Spanish MICINN's Consolider-Ingenio 2010 Programme under grant Multi-Dark 
{\bf CSD2009-00064}, the contract {\bf FPA2010-17747}, the France-US PICS no. 06482 and the LIA-TCAP of CNRS. 
Y.~M. acknowledges partial support from the ERC advanced grants Higgs@LHC and MassTeV and from  the European Union's Horizon 2020 research and innovation (programme under the Marie Sklodowska-Curie grant agreements No 690575 and No 674896).
This research was also supported in part by the Research Executive Agency (REA) of the European Union under
the Grant Agreement {\bf PITN-GA2012-316704} (``HiggsTools'') and by the CEFIPRA Project No. 5404-2: "Glimpses of New Physics".

\appendix


\section{Decay widths of the $Z^\prime$ boson}

\subsection{Kinetic mixing scenario}
\begin{equation}
\Gamma_{XX}=\frac{(g^{Z^\prime }_{X})^2 \sqrt{\frac{m_{Z'}^2}{4}-m_X^2} \left(m_{Z'}^2-4 m_X^2\right)^2}{48 \pi  m_X^2 m_{Z'}^2}.
\end{equation}
\begin{equation}
\Gamma_{Zh}=\frac{(g_{hZZ^\prime })^2 \Big(-2 m_{Z'}^2 \left(m_h^2-5 m_Z^2\right)+\left(m_h^2-m_Z^2\right)2+m_{Z'}^4\Big) \sqrt{\frac{\left(-m_h^2+m_Z^2+m_{Z'}^2\right)^2}{4 m_{Z'}^2}-m_z^2}}{96 \pi  m_Z^2 m_{Z'}^4}.
\end{equation}

\beq
\Gamma_{\ovl f f}= n_c \frac{ \sqrt{\frac{m_{Z'}^2}{4}-m_f^2} \Big[m_f^2 \Big(2 (v_f^{Z'})^2-4 (a_f^{Z'})^2\Big)+m_{Z^\prime}^2 \Big((a_f^{Z'})^2+(v_f^{Z'})^2\Big)\Big]}{6 \pi  m_{Z^\prime}^2},
\eeq
where $n_c$ denotes the colour factor, equals to 3 for quarks and 1 for leptons. $a_f^{Z'}=\dfrac{g^{Z^\prime}_{f_R}-g^{Z^\prime}_{f_L}}{2}$ and $v_f^{Z'}=\dfrac{g^{Z^\prime}_{f_R}+g^{Z^\prime}_{f_L}}{2}$ are the axial and vector couplings respectively.

\begin{equation}
\Gamma_{W^+W^-}=\frac{(g^{Z^\prime }_{W})^2 \sqrt{\frac{m_{Z^\prime}^2}{4}-m_W^2} \left(-68 m_W^4 m_{Z^\prime}^2+16 m_W^2 m_{Z^\prime}^4-48 m_W^6+m_{Z^\prime}^6\right)}{96 \pi  m_W^4 m_{Z^\prime}^2}.
\end{equation}

\subsection{Two Chern-Simons couplings scenario}
\begin{equation}
\Gamma_{XX}=\frac{\alpha_\text{CS} ^2 \left(m_{Z^\prime}^2-4 m_X^2\right)^2 \sqrt{m_{Z^\prime}^4-4 m_X^2 m_{Z^\prime}^2}}{24 \pi m_X^2 m_{Z^\prime}^3}.
\end{equation}
\begin{equation}
\Gamma_{Z\gamma}=\frac{\beta_\text{CS}^2 \cos^2\theta_W \left(m_{Z^\prime}^2-m_Z^2\right)^3 \left(m_{Z^\prime}^2+m_Z^2\right)}{24 \pi  m_z^2 m_{Z^\prime}^5}.
\end{equation}
\beq
\Gamma_{ZZ}= \frac{\beta_{\rm CS} ^2 \sin^2\theta_W \left(m_{Z^\prime}^2-4 m_Z^2\right)^2 
\sqrt{m_{Z^\prime}^2-4 m_Z^2 }}{24 \pi  m_Z^2 m_{Z^\prime}^2}.
\eeq

\section{Computation of the Chern-Simons coupling}
\label{sec:appendixC}

In this section of the appendix we show how to derive an effective Lagrangian from a UV complete model framework, as considered in this work. 
To achieve this goal, we review computations performed in ref.~\cite{Anastasopoulos:2006cz} of potential anomalous diagrams of the theory which involve three external gauge 
bosons or axions, interacting through a triangular loop of massive fermions. We can classify the possible diagrams in three categories: 
(1) diagrams without any mass insertions. Such diagrams are linearly divergent and proportional to the usual anomaly 
trace~\cite{Adler:1969gk,Bell:1969ts} (see ref. \cite{Bilal:2008qx} also for an overview). (2) 
Diagrams involving three gauge bosons with two mass insertions which give a finite result.
They are connected to the so-called  "CS" interaction (see, for example ref.~\cite{Anastasopoulos:2006cz})
and, (3) diagrams involving axions and 
two gauge bosons with one mass insertion. Just like the former
class of diagrams, these diagrams are also finite.
We will consider examples where the incoming state for the triangle loop
diagrams with heavy fermions is either a gauge boson $A^\mu_i(\vec{k}_3)$
or an axion $\theta_i$ while two outgoing states are two gauge fields,
denoted as $A^\rho_k(\vec{k}_2)$ and $A^\nu_j(\vec{k}_1)$, respectively. 
After evaluating these loop diagrams, we also compute the gauge transformations of the effective 
Lagrangian to ensure an anomaly free setup.

\subsection{Diagrams with two mass insertions: "Chern-Simons" contribution}

\begin{figure}[t!]
\centering
\includegraphics[width=7.5cm]{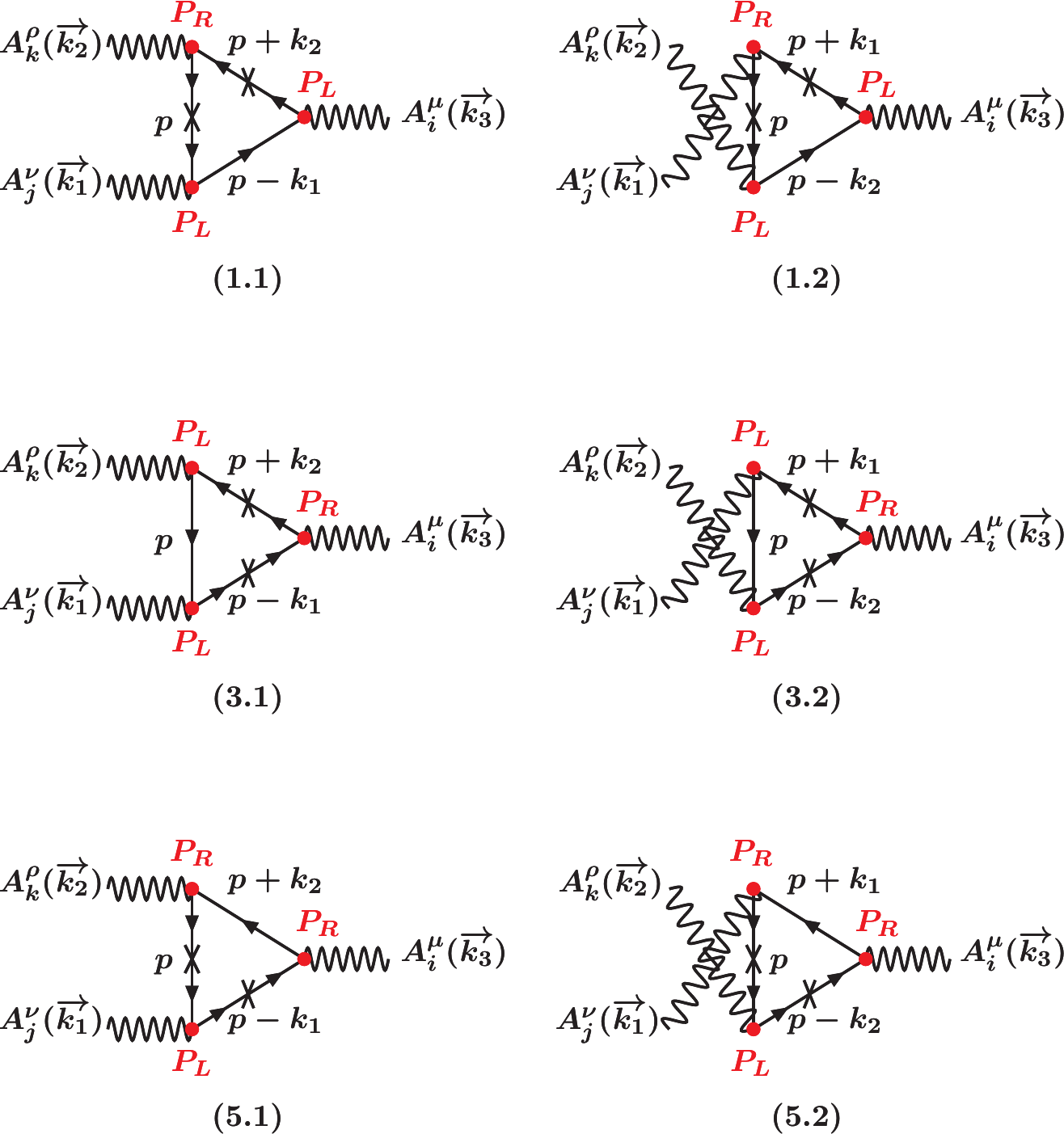}
\includegraphics[width=7.5cm]{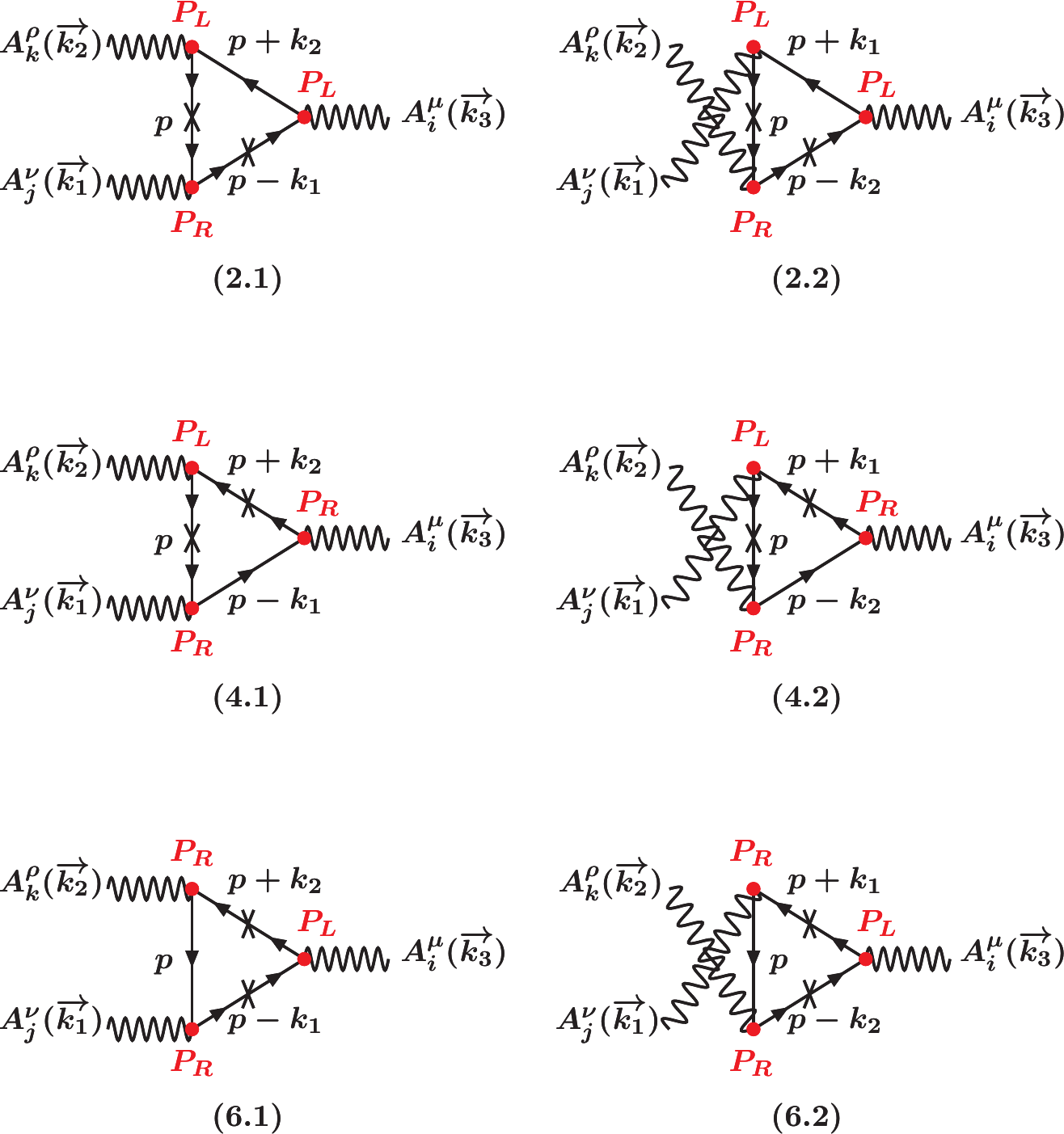}
\caption{Possible set of diagrams for triangular loops of heavy BSM fermions giving
rise to an effective CS interaction. Here {\boldmath `$\times$'} denotes
a mass insertion and the arrows represent the direction of fermion flow. Directions 
of the momentum, i.e., $p,\,p\pm k_1,\,p\mp k_2$, etc. are opposite to the fermion
flow.}
\label{fig:CSgenerate}
\end{figure}

We initiate our analysis for diagrams having gauge fields in the three external legs
and two mass insertions, which is equivalent to have two chirality flips for each diagram. Thus, we can chose one 
dominant chirality over the remaining two others in the fermionic loop and have three  possibilities to place the 
mass insertions for each dominant chirality. Further, considering the fact
that we can contract the external legs of the 
outgoing gauge fields in two different ways, we end up with twelve
diagrams as shown in figure \ref{fig:CSgenerate}. 

For a systematic analysis we start with the following Lagrangian:
%
\begin{equation}
\mathcal{L}\supset \sum_{a=i,j,k} i\ovl{F}\partial_\mu \gamma^\mu F-m\ovl{F}F-Q^a_L 
\ovl{F}_L\gamma_\mu F_L A^\mu_a-Q^a_R \ovl{F}_R\gamma_\mu F_R A^\mu_a,
\end{equation} 
where $F=F_L+F_R$ are the BSM heavy fermions running in the triangle
loops with $Q^a_L,\,Q^a_R$ as the relevant gauge charges
associated with $A^a_\mu$ field for the left- and right-chiral BSM fermions, respectively.

We can write contributions of the twelve diagrams without contractions on 
any external legs as an integral over the momentum $p$ as:
\begin{equation}
\label{eq:gammafracCS}
\int \dfrac{d^4p}{(2\pi)^4}\Gamma^{\mu \nu \rho}(p,k_1,k_2),
\end{equation}
where $\Gamma^{\mu \nu \rho}$, after integrating out heavy fermionic
degrees of freedom, can be expanded in the powers of external
momentums to achieve the low-energy effective Lagrangian. The 
expansion goes as:
\bea
\label{eq:Fullgammapart}
\Gamma^{\mu \nu \rho}(p,k_1,k_2)&&\simeq \Gamma^{\mu \nu \rho}(p,0,0)
+\sum^{2}_{i=1} k_i^\alpha \left.\left(\dfrac{\partial \Gamma^{\mu \nu \rho}(p,k_1,k_2)}{\partial k_i^\alpha}\right)\right\vert_{k_i=0}
\nonumber\\
&&+\frac{1}{2!} \sum^{2}_{i,j=1} k_i^\alpha k_j^\beta 
\left.\left(\dfrac{\partial^2 \Gamma^{\mu \nu \rho}(p,k_1,k_2)}{\partial k_i^\alpha \partial k_j^\beta}\right)\right\vert_{k_{i,j}=0}
+ \mathcal{O} (k^3_{i,\,j,\,k}).  
\eea

Now from figure \ref{fig:CSgenerate} it is apparent that $\Gamma^{\mu \nu \rho}(p,k_1,k_2)$
can be decomposed as a product of the two terms,
i.e., $\Gamma^{\mu \nu \rho}(p,k_1,k_2)=\bold{\Pi} \cdot \bold{TR}^{\mu \nu \rho}$ where $\bold{TR}^{\mu \nu \rho}$ includes 
possible couplings and traces over gamma matrices 
while $\bold{\Pi}$ is defined in the following way:
\bea
\label{eq:picalc}
\bold{\Pi}&=&\dfrac{1}{p^2-m^2}\dfrac{1}{(p+k_2)^2-m^2}\dfrac{1}{(p-k_1)^2-m^2} \nonumber\\
&& \hspace*{1cm}\text{for diagrams
like x.1 with x=1,2,...,6}, \nonumber\\ 
{\rm and}~\bold{\Pi}&=&\dfrac{1}{p^2-m^2}\dfrac{1}{(p-k_2)^2-m^2}\dfrac{1}{(p+k_1)^2-m^2} \nonumber\\
&& \hspace*{1cm}\text{for diagrams 
like x.2 with x=1,2,...,6}.~~~~ 
\eea

The trace in the leading term $\Gamma(p,0,0)$ 
(see eq.~(\ref{eq:Fullgammapart}))
appears to be proportional to odd powers of $p$ for  the 
numerator, which vanishes after $\int d^4p$ integration. 
The linear and quadratic terms in $k_i^\alpha$ can be computed 
straightforwardly from eq.~(\ref{eq:Fullgammapart}) as :
\bea
\label{eq:kalphaexp}
k_i^\alpha \left. \left( \dfrac{\partial \Gamma^{\mu \nu \rho}(p,k_1,k_2)}{\partial k_i^\alpha}  
\right) \right\vert_{k_i=0} &&=k_i^\alpha \left[ \left( \dfrac{\partial \bold{TR}^{\mu \nu \rho}}{\partial k_i^\alpha} 
\cdot \bold{\Pi} \right)  + \left( \dfrac{\partial \bold{\Pi}}{\partial k_i^\alpha} 
\cdot \bold{TR}^{\mu \nu \rho} \right) \right]_{k_i=0},\nonumber\\
k_i^\alpha k_j^\beta \left. \left( \dfrac{\partial^2 \Gamma^{\mu \nu \rho}(p,k_1,k_2)}
{\partial k_i^\alpha \partial k_j^\beta} \right) \right\vert_{k_{i,j}=0} &&=
k_i^\alpha k_j^\beta \left[ \left( \dfrac{\partial^2 \bold{TR}^{\mu \nu \rho}}{\partial k_i^\alpha \partial k_j^\beta} 
\cdot \bold{\Pi} \right)  +  
 \left( \dfrac{\partial^2 \bold{\Pi}}{\partial k_i^\alpha \partial k_j^\beta} 
\cdot \bold{TR}^{\mu \nu \rho} \right) \right.\nonumber\\
&&\left. +
 \left( \dfrac{\partial \bold{TR}^{\mu \nu \rho}}{\partial k_i^\alpha} 
\cdot \dfrac{\partial \bold{\Pi}}{\partial k_j^\beta}  \right)  + 
 \left( \dfrac{\partial \bold{TR}^{\mu \nu \rho}}{\partial k_j^\beta} 
\cdot \dfrac{\partial \bold{\Pi}}{\partial k_i^\alpha}  \right) \right]_{k_{i,j}=0}.~~~~
\eea

The contributions from denominators $\bold{\Pi}$ are shown
in eq.~(\ref{eq:picalc}) while trace contributions 
$\bold{TR^{\mu\nu\rho}}$ from the twelve diagrams are:
\bea
\label{eq:trcalccs}
\bold{TR}^{\mu \nu \rho}_{1.1}&&=Q_L^iQ_L^jQ_R^k\text{Tr}[(\cancel{p}+m)\gamma^\rho P_R (\cancel{p}+\cancel{k}_2+m)\gamma^\mu P_L (\cancel{p}-\cancel{k}_1)\gamma^\nu P_L],\nonumber\\
\bold{TR}^{\mu \nu \rho}_{1.2}&&=Q_L^iQ_R^jQ_L^k\text{Tr}[(\cancel{p}+m)\gamma^\nu P_R (\cancel{p}+\cancel{k}_1+m)\gamma^\mu P_L (\cancel{p}-\cancel{k}_2)\gamma^\rho P_L],\nonumber\\
\bold{TR}^{\mu \nu \rho}_{2.1}&&=Q_L^iQ_R^jQ_L^k\text{Tr}[(\cancel{p}+m)\gamma^\rho P_L (\cancel{p}+\cancel{k}_2)\gamma^\mu P_L (\cancel{p}-\cancel{k}_1+m)\gamma^\nu P_R],\nonumber\\
\bold{TR}^{\mu \nu \rho}_{2.2}&&=Q_L^iQ_L^jQ_R^k\text{Tr}[(\cancel{p}+m)\gamma^\nu P_L (\cancel{p}+\cancel{k}_1)\gamma^\mu P_L (\cancel{p}-\cancel{k}_2+m)\gamma^\rho P_R],\nonumber\\
\bold{TR}^{\mu \nu \rho}_{3.1}&&=Q_R^iQ_L^jQ_L^k\text{Tr}[\cancel{p}\,\gamma^\rho P_L (\cancel{p}+\cancel{k}_2+m)\gamma^\mu P_R (\cancel{p}-\cancel{k}_1+m)\gamma^\nu P_L],\nonumber\\
\bold{TR}^{\mu \nu \rho}_{3.2}&&=Q_R^iQ_L^jQ_L^k\text{Tr}[\cancel{p}\,\gamma^\nu P_L (\cancel{p}+\cancel{k}_1+m)\gamma^\mu P_R (\cancel{p}-\cancel{k}_2+m)\gamma^\rho P_L],\nonumber\\
\bold{TR}^{\mu \nu \rho}_{4.1}&&=Q_R^iQ_R^jQ_L^k\text{Tr}[(\cancel{p}+m)\gamma^\rho P_L (\cancel{p}+\cancel{k}_2+m)\gamma^\mu P_R (\cancel{p}-\cancel{k}_1)\gamma^\nu P_R],\nonumber\\
\bold{TR}^{\mu \nu \rho}_{4.2}&&=Q_R^iQ_L^jQ_R^k\text{Tr}[(\cancel{p}+m)\gamma^\nu P_L (\cancel{p}+\cancel{k}_1+m)\gamma^\mu P_R (\cancel{p}-\cancel{k}_2)\gamma^\rho P_R],\nonumber\\
\bold{TR}^{\mu \nu \rho}_{5.1}&&=Q_R^iQ_L^jQ_R^k\text{Tr}[(\cancel{p}+m)\gamma^\rho P_R (\cancel{p}+\cancel{k}_2)\gamma^\mu P_R (\cancel{p}-\cancel{k}_1+m)\gamma^\nu P_L],\nonumber\\
\bold{TR}^{\mu \nu \rho}_{5.2}&&=Q_R^iQ_R^jQ_L^k\text{Tr}[(\cancel{p}+m)\gamma^\nu P_R (\cancel{p}+\cancel{k}_1)\gamma^\mu P_R (\cancel{p}-\cancel{k}_2+m)\gamma^\rho P_L],\nonumber\\
\bold{TR}^{\mu \nu \rho}_{6.1}&&=Q_L^iQ_R^jQ_R^k\text{Tr}[\cancel{p}\,\gamma^\rho P_R (\cancel{p}+\cancel{k}_2+m)\gamma^\mu P_L (\cancel{p}-\cancel{k}_1+m)\gamma^\nu P_R],\nonumber\\
\bold{TR}^{\mu \nu \rho}_{6.2}&&=Q_L^iQ_R^jQ_R^k\text{Tr}[\cancel{p}\,\gamma^\nu P_R (\cancel{p}+\cancel{k}_1+m)\gamma^\mu P_L (\cancel{p}-\cancel{k}_2+m)\gamma^\rho P_R],
\eea
where mass insertions are properly taken into account.

One can use eq.~(\ref{eq:kalphaexp}) to extract contributions
from the twelve diagrams shown in figure \ref{fig:CSgenerate}.
For example, the contribution proportional to $Q_L^iQ_L^jQ_R^k$ involving diagrams $(1.1)$ and $(2.2)$, in the linear 
terms of eq.~(\ref{eq:kalphaexp}), gives:
\bea
\int \dfrac{d^4p}{(2\pi)^4}\Big( \Gamma^{\mu \nu \rho}_{(1.1)}+\Gamma^{\mu \nu \rho}_{(2.2)} 
\Big)&&= 8im^2 Q_L^iQ_L^jQ_R^k \epsilon^{\mu \nu \rho \sigma} \nonumber\\
&& \times \int \dfrac{d^4p}{(2\pi)^4} 
\left[ \dfrac{(k_2-k_1)_\sigma}{4} \dfrac{p^2}{(p^2-m^2)^4} 
+\dfrac{k_{1 \sigma}}{2(p^2-m^2)^3} \right],\nonumber\\ 
&&=  Q_L^iQ_L^jQ_R^k \epsilon^{\mu \nu \rho \sigma} 
\dfrac{1}{24\pi^2}(k_3+k_1)_\sigma,
\eea
where we used the known expressions for different momentum
integrals over $p$ (see ref.~\cite{Peskin:1995ev} for example) 
and $k_1+k_2=k_3$.
In a similar way the contributions from all the twelve
diagrams of figure \ref{fig:CSgenerate} can be grouped
as shown in table \ref{tab:CScont}.
\begin{table}[t!]
\begin{center}
\label{tab:CScont}
\begin{tabular}{|c||c|}
\hline 
Diagrams & Contribution to $\Gamma^{\mu \nu \rho}$ \\ 
\hline 
(1.1)+(2.2) & $Q_L^iQ_L^jQ_R^k \epsilon^{\mu \nu \rho \sigma} (k_3+k_1)_\sigma /(24\pi^2)$ \\ 
\hline
(2.1)+(1.2) & $-Q_L^iQ_R^jQ_L^k \epsilon^{\mu \nu \rho \sigma} (k_3+k_2)_\sigma/(24\pi^2)$ \\ 
\hline
(3.1)+(3.2) &  $Q_R^iQ_L^jQ_L^k \epsilon^{\mu \nu \rho \sigma} (k_2-k_1)_\sigma/(24\pi^2)$ \\ 
\hline
(4.1)+(5.2) &  $-Q_R^iQ_R^jQ_L^k \epsilon^{\mu \nu \rho \sigma} (k_3+k_1)_\sigma/(24\pi^2)$ \\
\hline
(5.1)+(4.2) &  $Q_R^iQ_L^jQ_R^k \epsilon^{\mu \nu \rho \sigma} (k_3+k_2)_\sigma/(24\pi^2)$ \\
\hline
(6.1)+(6.2) &  $Q_L^iQ_R^jQ_R^k \epsilon^{\mu \nu \rho \sigma} (k_1-k_2)_\sigma/(24\pi^2)$ \\  
\hline
\end{tabular} 
\end{center}
\caption{Resultant contributions of the twelve diagrams of figure
\ref{fig:CSgenerate}, clubbed according to the same pre-factor.}
\end{table}

From table \ref{tab:CScont}, one can factorize the sum of all contributions proportional to the external momentum $k_3$ as:

\bea
\int \dfrac{d^4p}{(2\pi)^4}\Gamma^{\mu \nu \rho} &&\supset \dfrac{1}{24\pi^2} \epsilon^{\mu \nu \rho \sigma} 
k_{3 \sigma} (Q_L^iQ_L^jQ_R^k-Q_L^iQ_R^jQ_L^k-Q_R^iQ_R^jQ_L^k+Q_R^iQ_L^jQ_R^k),\nonumber\\
&&\supset\dfrac{1}{24\pi^2} \epsilon^{\mu \nu \rho \sigma} k_{3 \sigma} (Q_L^i+Q_R^i)(Q_L^jQ_R^k-Q_L^kQ_R^j).
\eea
The same factorization can be done for $k_2$ and $k_1$
to produce the following effective Lagrangian :
\begin{equation}
\label{eq:LeffCS}
\mathcal{L}^\text{eff}_{\rm CS} = \dfrac{1}{96\pi^2} (Q_L^k+Q_R^k)(Q_L^iQ_R^j-Q_L^jQ_R^i) 
\epsilon_{\mu \nu \rho \sigma} A_i^\mu A_j^\nu F_k^{\rho \sigma},
\end{equation}
where summation over all the possible combinations of the 
gauge fields is implied. 

\subsection{Diagrams with axions}

\begin{figure}[t!]
\centering
\includegraphics[width=8.65cm]{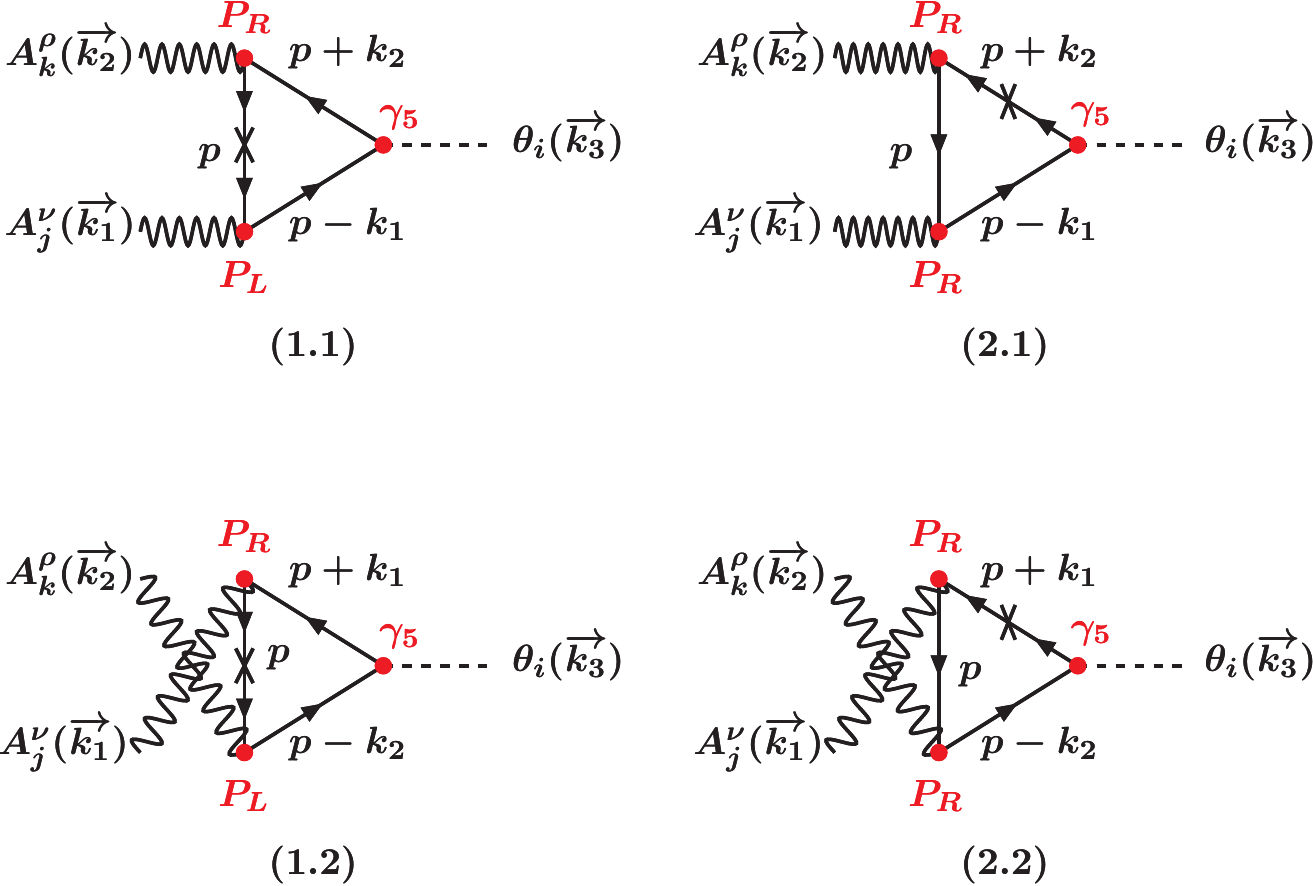}
\includegraphics[width=4cm]{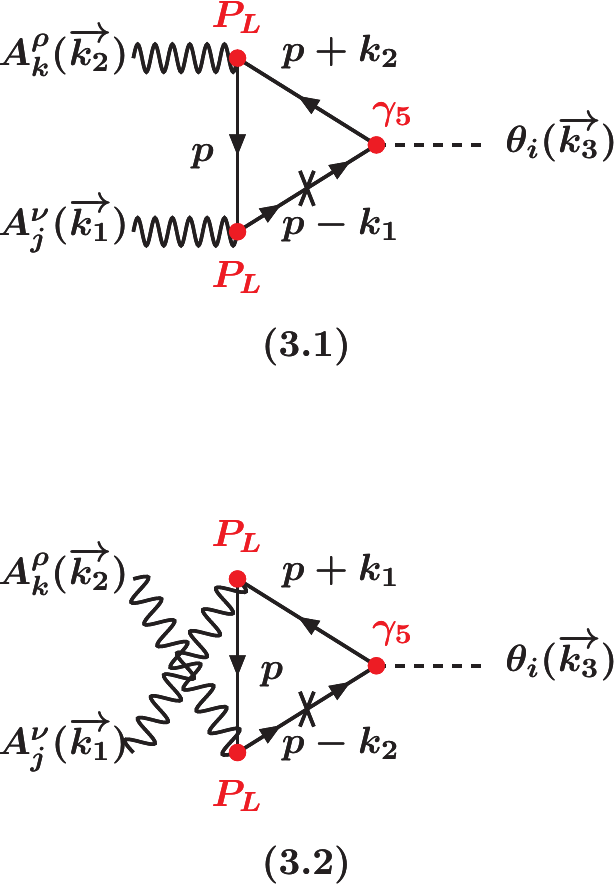}
\caption{Possible set of diagrams for triangular loops of heavy BSM fermions with
an external axion field. Here {\boldmath `$\times$'} denotes
a mass insertion and the arrows represent the direction of fermion flow. Directions 
of the momentum, i.e., $p,\,p\pm k_1,\,p\mp k_2$, etc. are opposite to the fermion
flow.}
\label{fig:CSaxion}
\end{figure}

The diagrams involving an axion field $\theta_i$ include only one mass insertion since vertices with two fermionic legs and an axion field
flips the chirality, as evidenced from the following Lagrangian:
\begin{equation}
\mathcal{L}^\text{axion}=-iy_F \theta_i \ovl{F}_L  F_R +\text{h.c.}-m\ovl{F}F
=-i y_F \theta_i \ovl{F}\gamma_5 F -m\ovl{F}F,
\end{equation}
where  $y_F$ is the associated Yukawa coupling. For the chosen Lagrangian we have three possible ways to place a mass insertion on the
fermionic propagators and two different ways to connect the external lines with the vertices, giving a total of six 
diagrams as shown in figure \ref{fig:CSaxion}.

Once again, like the CS case (see eq.~(\ref{eq:gammafracCS})),
we can write the sum over the
six diagrams without contractions on the external legs as an 
integral over the momentum $p$ as:
\begin{equation}
\int \dfrac{d^4p}{(2\pi)^4}\Gamma^{\nu \rho}(p,k_1,k_2),
\end{equation}
where $\Gamma^{\nu \rho}(p,k_1,k_2)$ is decomposed as
$\bold{\Pi}\cdot\bold{TR}^{\nu \rho}$ with $\bold{\Pi}$
as already defined in eq.~(\ref{eq:picalc}) and 
the trace factors $\bold{TR}^{\nu \rho}$ written as:
%
\bea
\label{eq:traceano}
\bold{TR}^{\nu \rho}_{1.1}&&=y_FQ_L^jQ_R^k  \text{Tr}[\gamma_5(\cancel{p}-\cancel{k}_1)\gamma^\nu 
P_L (\cancel{p} +m) \gamma^\rho P_R (\cancel{p}+\cancel{k}_2) ],\nonumber\\
\bold{TR}^{\nu \rho}_{1.2}&&=y_FQ_R^jQ_L^k  \text{Tr}[\gamma_5(\cancel{p}-\cancel{k}_2)\gamma^\rho P_L 
(\cancel{p} +m) \gamma^\nu P_R (\cancel{p}+\cancel{k}_1)],\nonumber\\
\bold{TR}^{\nu \rho}_{2.1}&&=y_F Q_R^jQ_R^k  \text{Tr}[\gamma_5(\cancel{p}-\cancel{k}_1)\gamma^\nu P_R 
\,\cancel{p}\,\gamma^\rho P_R (\cancel{p}+\cancel{k}_2+m)],\nonumber\\
\bold{TR}^{\nu \rho}_{2.2}&&=y_F Q_R^jQ_R^k  \text{Tr}[\gamma_5(\cancel{p}-\cancel{k}_2)\gamma^\rho P_R 
\,\cancel{p}\,\gamma^\nu P_R (\cancel{p}+\cancel{k}_1+m)],\nonumber\\
\bold{TR}^{\nu \rho}_{3.1}&&=y_F Q_L^jQ_L^k  \text{Tr}[\gamma_5
(\cancel{p}-\cancel{k}_1+m) \gamma^\nu P_L \,\cancel{p}\,\gamma^\rho P_L (\cancel{p}+\cancel{k}_2) ],\nonumber\\
\bold{TR}^{\nu \rho}_{3.2}&&=y_F Q_L^jQ_L^k  \text{Tr}[\gamma_5 (\cancel{p}-\cancel{k}_2+m)
\gamma^\rho P_L \,\cancel{p}\,\gamma^\nu P_L (\cancel{p}+\cancel{k}_1)],
\eea
with proper mass insertion.

Now if we consider expansion of $\Gamma^{\nu \rho}(p,k_1,k_2)$ 
in powers of the external momentums $k_1,\,k_2$, just like the already studied CS
scenario, the zeroth and linear order terms of the expansion vanish.
This happens as the former is proportional to $k_i$ while the latter
yields contributions $\propto p$ and thus, disappears after performing $\int d^4 p$
over an odd function. The leading contribution thus, comes from the second order. Considering a 
CP invariant UV complete theory, we keep only the CP-odd contribution in $\Gamma^{\nu \rho}(p,k_1,k_2)$ since axions are CP-odd fields. 
With this approach one can compute the detail expressions for all the six
diagrams of figure \ref{fig:CSaxion}, for example, for the diagram $(3.1)$ one gets:
\bea
\int \dfrac{d^4p}{(2\pi)^4}\Gamma^{\nu \rho}_{(3.1)}(p,k_1,k_2)&&= im y_F Q_L^j Q_L^k 
\epsilon^{\mu \nu \rho \sigma}k_{1 \rho} k_{2 \sigma} \int \dfrac{d^4p}{(2\pi)^4} \dfrac{p^2}{(p^2-m^2)^4},\nonumber\\
&&=\frac{1}{v_i}\frac{Q_L^jQ_L^k}{48\pi^2} \epsilon^{\nu \rho \alpha \beta} k_{1 \alpha} k_{2 \beta},
\eea
where in the last step we have used the known momentum integral as of ref. \cite{Peskin:1995ev}
as well as $m=y_F v_i$ with $v_i$ being the VEV of the 
scalar field giving masses to the BSM fermions $F$ and the gauge field $A^\mu_i$. In a similar way the contributions from all 
the six diagrams of
figure \ref{fig:CSaxion} can be evaluated as tabulated
in table \ref{tab:Axioncont}.
\begin{table}[t!]
\label{tab:Axioncont} 
\begin{center}
\begin{tabular}{|c||c|}
\hline 
Diagrams & Contribution to $\Gamma^{\nu \rho}$ \\ 
\hline 
(1.1) & $Q_L^jQ_R^k \epsilon^{\nu \rho \alpha \beta} k_{1 \alpha} k_{2 \beta}/(48\pi^2v_i)$ \\ 
\hline
(2.1) & $Q_R^jQ_R^k \epsilon^{\nu \rho \alpha \beta} k_{1 \alpha} k_{2 \beta}/(48\pi^2v_i)$ \\ 
\hline
(3.1) &  $Q_L^jQ_L^k \epsilon^{\nu \rho \alpha \beta} k_{1 \alpha} k_{2 \beta}/(48\pi^2v_i)$ \\ 
\hline
(1.2) &  $Q_R^jQ_L^k \epsilon^{\nu \rho \alpha \beta} k_{1 \alpha} k_{2 \beta}/(48\pi^2v_i)$ \\
\hline
(2.2) &  $Q_R^jQ_R^k \epsilon^{\nu \rho \alpha \beta} k_{1 \alpha} k_{2 \beta}/(48\pi^2v_i)$ \\
\hline
(3.2) &  $Q_L^jQ_L^k \epsilon^{\nu \rho \alpha \beta} k_{1 \alpha} k_{2 \beta}/(48\pi^2v_i)$ \\  
\hline
\end{tabular} 
\end{center}
\caption{Final contributions of the 6 diagrams of figure
\ref{fig:CSaxion}.}
\end{table}

Summing all the 6 contributions from table \ref{tab:Axioncont} yields :
\begin{equation}
\int \dfrac{d^4p}{(2\pi)^4}\Gamma^{ \nu \rho} = \dfrac{1}{v_i} \dfrac{1}{48\pi^2} 
\epsilon^{\nu \rho \alpha \beta} k_{1 \alpha} k_{2 \beta} [2(Q_L^jQ_L^k+Q_R^jQ_R^k )+Q_R^jQ_L^k +Q_L^jQ_R^k ],
\end{equation}
which finally produce the following effective Lagrangian:
\begin{equation}
\label{eq:Leffaxion}
\mathcal{L}^\text{eff}_{\rm axion} = \dfrac{1}{192\pi^2}[2(Q_L^jQ_L^k+Q_R^jQ_R^k )+Q_R^jQ_L^k +Q_L^jQ_R^k ] 
\epsilon_{\mu \nu \rho \sigma} \frac{\theta_i}{v_i} F_j^{\mu \nu} F_k^{\rho \sigma},
\end{equation}
where, once again summation over all the possible combinations of the 
gauge fields is implied.

\subsection{Anomaly cancellation}
In the last two subsections we discussed about the three different anomalous
contributions, namely, (1) diagrams without any chirality flip which
are linearly divergent and giving contributions proportional to the anomaly
traces. (2) Diagrams with one chirality flip involving an axion field
that are finite and, (3) the so-called "CS" contributions
which are finite and invoke two chirality flips. In this subsection
we show that gauge transformation of the effective CS
and axion Lagrangians (see eq.~(\ref{eq:LeffCS}) and eq.~(\ref{eq:Leffaxion}))
is proportional to the ``usual'' anomaly trace. Hence, a vanishing
anomaly trace, with appropriate distribution of the charges of BSM fermions,
assures an anomaly free theory construction. Given the following
gauge transformations of an axion field $\theta_i$ and a gauge field
$A^\mu_i$: 
\beq
\label{eq:CSaxionGT}
\theta_i  \rightarrow \theta_i + v_i (Q_L^i-Q_R^i) \alpha_i, \,\,\,\,
A^\mu_i  \rightarrow A^\mu_i + \partial^\mu \alpha_i,
\eeq
where $\alpha_i$ is the parameter of gauge transformation,
the variation of the effective CS Lagrangian
(see eq.~(\ref{eq:LeffCS})) becomes:
\begin{equation}
\label{eq:LCSvarn}
\delta\mathcal{L}^\text{CS} =  -\dfrac{1}{192\pi^2} \Big[(Q_L^k+Q_R^k)(Q_L^iQ_R^j-Q_L^jQ_R^i)
+(Q_L^j+Q_R^j)(Q_L^iQ_R^k-Q_L^kQ_R^i)\Big] \epsilon_{\mu \nu \rho \sigma} \alpha_i F_j^{\mu \nu}  F_k^{\rho \sigma},
\end{equation}
where we have used the advantages of integrating by parts as well
as Bianchi identity and included all possible combinations of the $i,\,j,\,k$ indices. The change in effective axion Lagrangian (see eq.~(\ref{eq:Leffaxion})) is given by:
\begin{equation}
\label{eq:Laxionvarn}
\delta \mathcal{L}^\text{axion} = \dfrac{1}{192\pi^2}[2(Q_L^jQ_L^k+Q_R^jQ_R^k )+Q_R^jQ_L^k 
+Q_L^jQ_R^k ]  (Q_L^i-Q_R^i) \alpha_i \epsilon_{\mu \nu \rho \sigma} F_j^{\mu \nu} F_k^{\rho \sigma}.
\end{equation}

Combining eq.~(\ref{eq:LCSvarn}) and eq.~(\ref{eq:Laxionvarn})
the resultant variation, given the transformations of 
eq.~(\ref{eq:CSaxionGT}), is written as:
\begin{equation}
\label{eq:LCSaxiontotvarn}
\delta \mathcal{L}  = \delta \mathcal{L}^\text{axion} + \delta \mathcal{L}^\text{CS} = 
\dfrac{1}{96\pi^2} \alpha_i \Big[Q_L^iQ_L^jQ_L^k-Q_R^iQ_R^jQ_R^k \Big] 
\epsilon_{\mu \nu \rho \sigma} F_j^{\mu \nu} F_k^{\rho \sigma}.
\end{equation}

It is now apparent from eq. (\ref{eq:LCSaxiontotvarn}) that
gauge transformation of the total Lagrangian, i.e., axion and
CS Lagrangians is
proportional to the ``usual'' anomaly trace $Q_L^iQ_L^jQ_L^k-Q_R^iQ_R^jQ_R^k $. 
Hence, with proper choice of the charges for the heavy fermions
one can ensure an anomaly free theory setup where the anomaly trace
$Q_L^iQ_L^jQ_L^k-Q_R^iQ_R^jQ_R^k $ vanishes for all $i,j,k$.
Further, when this anomaly trace disappears with proper choice of $Q^i_L,\,Q^i_R,\,Q^j_L,\,Q^j_R,\,Q^k_L$ and $Q^k_R$,
the combination of the axion and CS effective Lagrangians
(i.e., eq.~(\ref{eq:LeffCS}) + eq.~(\ref{eq:Leffaxion})) can be 
embedded into a dimension-six operator as:
\begin{equation}
\epsilon_{\mu \nu \rho \sigma} D^\mu \theta_i D^\nu \theta_j F_k^{\rho \sigma},
\end{equation}
where $D^\mu,\,D^\nu$ are co-variant derivatives for 
axion fields $\theta_i,\,\theta_j$, respectively. One can always
consider the case of unitary gauge when the axion Lagrangian
(see eq.~(\ref{eq:Leffaxion})) vanishes and the total Lagrangian
is simply the CS one, as given by eq.~(\ref{eq:LeffCS}).
This is the scenario which we studied in this work. Recasting
eq.~(\ref{eq:LeffCS}) for the specific case of $U(1)_X\times U(1)_V$,
as considered in this work, one can generate
%
\begin{equation}
\mathcal{L} = \mathcal{L}^{\rm eff}_{\rm CS} = \dfrac{1}{48\pi^2} (Q_L^X+Q_R^X)(Q_L^XQ_R^V-Q_L^VQ_R^X) 
\epsilon_{\mu \nu \rho \sigma} X^\mu \wt V^{\nu} X^{\rho \sigma} \equiv 
\alpha_\text{CS} \epsilon_{\mu \nu \rho \sigma} X^\mu \wt V^{\nu}  X^{\rho \sigma},
\end{equation}
where terms $\propto \epsilon_{\mu \nu \rho \sigma} \wt V^\mu X^{\nu}  \wt V^{\rho \sigma}$
is effaced with suitable choice of associated charges and the parameter
$\alpha_{\rm CS}$ is given by
\begin{equation}
\label{eq:effctivealphaCS}
\alpha_\text{CS} \equiv \dfrac{1}{48\pi^2} (Q_L^X+Q_R^X)(Q_L^XQ_R^V-Q_L^VQ_R^X),
\end{equation}
which we have already used to derive eq.~(\ref{eq:alphaCSwork})
for the charges given in table \ref{tab:charges}.

\bibliography{bibfile2}{}

\end{document}